\documentclass[useAMS,usenatbib]{mn2e}

\usepackage{amsmath,amssymb}
\usepackage{graphicx}
\usepackage{epstopdf}
\usepackage{color}
\usepackage[normalem]{ulem}
\usepackage[usenames,dvipsnames]{xcolor}
\usepackage{aas_macros}
\usepackage{url}

\newcommand{\lal}{Lyman-$\alpha$ }
\newcommand{\lab}{Lyman-$\beta$ }

\newcommand{\ulas}{ULAS J1120+0641 }

\newcommand{\oone}{O~{\small I} }
\newcommand{\cfour}{C~{\small IV} }

\newcommand{\magtwo}{Mg~{\small II} }

\newcommand{\red}[1]{\textcolor{black}{#1}}

\author[S. E. I. Bosman et al.]
  {Sarah E. I. Bosman$^{1,2,7}$\thanks{s.bosman@ucl.ac.uk}, 
Xiaohui Fan$^{3}$, Linhua Jiang$^{4}$, 
Sophie Reed$^{1}$, 
\newauthor Yoshiki Matsuoka$^{5}$, George Becker$^{6}$ \& Martin Haehnelt$^{1,2}$\\
  $^1$Institute of Astronomy, University of Cambridge, Madingley Road,
Cambridge CB3 0HA, U.K.\\
  $^2$Kavli Institute for Cosmology, University of Cambridge, Madingley Road,
Cambridge CB3 0HA, U.K. \\
  $^3$Steward Observatory, University of Arizona, Tucson, AZ 85721-0065, USA\\
  $^4$Kavli Institute for Astronomy and Astrophysics, Peking University, Beijing 100871, China\\
  $^5$Research Center for Space and Cosmic Evolution, Ehime University, Matsuyama, Ehime 790-8577, Japan\\
  $^6$Department of Physics \& Astronomy, University of California, Riverside, 900 University Avenue, Riverside, CA, most likely in92521, USA \\
  $^7$Department of Physics and Astronomy, University College London, London, WC1E 6BT, UK} 

\title{New constraints on \lal opacity with a sample of 62 quasars at $z>5.7$}
\date{}
\begin{document}
\maketitle

\begin{abstract}

We present measurements of the mean and scatter of the IGM \lal opacity at $4.9<z<6.1$ along the lines of sight of 62 quasars at $z_\text{source} > 5.7$, the largest sample assembled at these redshifts to date by a factor of two. The sample size enables us to sample cosmic variance at these redshifts more robustly than ever before. The spectra used here were obtained by the SDSS, DES-VHS and SHELLQs collaborations, drawn from the ESI and X-Shooter archives, reused from previous studies or observed specifically for this work. We measure the effective optical depth of \lal in bins of 10, 30, 50 and 70 cMpc $\text{h}^{-1}$, construct cumulative distribution functions under two treatments of  upper limits on flux and explore an empirical analytic fit to residual \lal transmission. We verify the consistency of our results with those of previous studies via bootstrap re-sampling and confirm the existence of tails towards high values in the opacity distributions, which may persist down to $z\sim5.2$. Comparing our results with predictions from cosmological simulations, we find further strong evidence against models that include a spatially uniform ionizing background and temperature-density relation. We also compare to IGM models that include either a fluctuating UVB dominated by rare quasars or temperature fluctuations due to patchy reionization.  Although both models produce better agreement with the observations, neither fully captures the observed scatter in IGM opacity.
Our sample of 62 $z>5.7$ quasar spectra opens many avenues for future study of the reionisation epoch.

\end{abstract}

\begin{keywords}
dark ages, reionization, first stars\  \  quasars: absorption lines\   \   intergalactic medium 
\end{keywords}

\section{Introduction} 

The first billion years of the Universe are currently a frontier of late-time cosmology, both observationally and theoretically. 
During this stretch of time the first stars and galaxies assembled from the primordial gas left behind by reheating, and the atomic hydrogen permeating the early Universe became ionised.
This ``reionization'' transition is believed to be largely completed by redshift six.
The precise timing and topology of reionisation are strongly influenced by the processes at work in the first galaxies and active galactic nuclei (AGN), as well as the large--scale structure of the early intergalactic medium (IGM). 

Quasars located at $z\gtrsim6.0$ have proven to be useful tools for obtaining information about reionisation due to their high intrinsic luminosities and prominent \lal emission lines. These properties have yielded results on multiple fronts, from measuring the sizes of quasar proximity zones across time, which are expected to diminish with increasing IGM neutrality and decreasing quasar lifetime (e.g. \citealp{Fan06,Carilli10,Keating15}) to  constraining  enrichment processes by probing the cosmic abundances of intervening metals (e.g. \citealp{RyanWeber09, Dodorico13, Beckerreview, Chen16, Bosman17}).
The \lal forest extending bluewards of $1215$\AA \ in the quasar rest frame is of particular interest to reionisation as it traces the diffuse intergalactic gas whose ionization is sensitive to the metagalactic ultraviolet background (UVB). 
The \lal opacity in the forest increases with redshift, and eventually complete absorption is reached once the IGM reaches average hydrogen neutral fractions of $f \geq 0.1$ per cent \citep{GP}. 
The characterisation of \lal opacity across redshift is a powerful constraint on models of reionisation, as the amount of residual transmission is sensitive to the nature of the UV sources, the thermal state of the IGM, and the large-scale clustering of sources among other factors (e.g. \citealt{Wyithe11,McQuinn11,Davies17}).

\lal transmission along QSO lines of sight is often quantified by an ``effective optical depth", $\tau = - \text{ln}(\langle F /F_0 \rangle)$, where $ F $ is the observed (residual) flux in the \lal forest, and $F_0$ is the unabsorbed continuum (e.g. \citealt{Fan06}). 
The first studies of the optical depth distribution pointed to a large scatter in transmission along lines of sight \citep{Songaila04,Fan06}.  Although this scatter was potentially incompatible with the predicted scatter due to large-scale fluctuations in the density field alone \citep{Lidz06}, firm conclusions were limited by the relatively modest sample sizes (e.g. \citealt{Mesinger10}).
\citet{Becker15} discovered a $\sim110$ cMpc \lal trough extending down to $z\simeq5.5$ and demonstrated that its existence, as well as the general distribution of \lal opacity measurements at $z \gtrsim 5.6$, is incompatible with a spatially uniform UVB. The discovery prompted a flurry of new reionisation models (see e.g. \citealt{Chardin15,Daloisio15,Davies16}). The combined samples of \citet{Fan06} and \citet{Becker15} included 26 quasars at $z > 5.7$, which is  only a fraction of the more than 200 quasars now known at these redshifts. In this paper we gather 62 spectra of $z>5.7$ quasars, more than doubling previous samples.

We present updated measurements of the \lal opacity distribution function (PDF) for the redshift range $4.9<z<6.1$. The number of known quasars at $z>5.9$ is increasing rapidly due to searches by the Dark Energy Survey (DES; \citealp{Reed15}), the  Subaru High-$z$ Exploration of Low-Luminosity Quasars (SHELLQs; \citealp{Matsuoka16}), Pan-STARRS \citep{Kaiser10}, the VISTA Kilo-Degree Infrared Galaxy (VIKING; \citealt{Venemans13,Carnall15}) survey, the Canada-France High-redshift Quasar Survey (CFHQS; \citealp{Willott07}) and UKIDSS (\citealt{Venemans07,Mortlock09,Mortlock11}) as well as the completion of the  search for high redshift quasars in the Sloan Digital Sky Survey (SDSS, \citealp{York2000}, \citealp{Jiang16}). 
Here we take advantage of this increase to provide smoother constraints on the \lal PDF with a better handle on cosmic variance. 
In addition, we are able to robustly sample the Lyman-alpha opacity distribution up to $z=6.1$ for the first time.

The paper is structured as follows. In Section 2 we describe our sample of 62
quasars and present four
previously unpublished spectra, briefly discussing the properties of our sample. Our methods for measuring \lal opacity distributions are presented, and compared to previous studies in Section 3. Challenges in dealing with the wide range of spectral resolutions and signal to noise ratios (SNRs) across our sample are discussed.
Section 4 gives our results spanning the redshift range $4.9<z<6.1$ using multiple ways of accounting for the inhomogeneous quality of the data and non-detections of transmitted flux. These results are confronted with predictions from numerical models and discussed in Section 5. 
Section 6 introduces our empirical functional form to residual \lal transmission and outlines our maximum likelihood fitting method.
We discuss implications for the process of reionisation and caveats of the work in Section 7.

The results are summarized in Section 8 and extra figures, including a mosaic of the entire sample, are shown in the Appendix. 
Throughout the paper we use 
$(\Omega m, \Omega \Lambda, h) = (0.308, 0.692, 0.678)$ \citep{Planck} and quote comoving distances in units of $\text{Mpc h}^{-1}$.
\red{We explicitly distinguish in all cases between $\tau$ and $\tau_\text{eff}$. The latter always refers to the \textit{binned} opacity, $\tau_\text{eff} = -\log (\langle F\rangle)$, as defined in more detail in Section 3. The binning scale is 50 cMpc h${}^{-1}$ everywhere except in Section 4.2, where we explicitly experiment with varying it.}
All measurements of $\tau_\text{eff}$ obtained and used in this paper are made available online\footnote{\red{\url{http://www.homepages.ucl.ac.uk/~ucapeib/data.html}}}.

\section{Data}

Our sample consists spectra of 62 quasars at $z > 5.7$ observed over the last 11 years. Out of these objects, 4 are discovery spectra from the SHELLQs survey, 10 were discovered by DES-VHS (out of which 4 are currently unpublished), 13 are SDSS discovery spectra, 13 are new reductions of archival data, 19 are adopted from previous studies on \lal transmission, and 3 are new to this work. Ten different optical spectrographs were used to obtain the data: ESI, X-Shooter, GMOS, MagE, EFOSC, FOCAS, MMT RCS, HIRES, MIKE and LBT-MODS. The following sections describe the make-up of the sample in more detail. Table~1 details the provenance of each spectrum. A mosaic of the entire sample is plotted in Appendix A.

\begin{table*}
\resizebox{\textwidth}{!}{
\vspace{-10em}
\begin{tabular}{l c c c c c c c}
\hline
QSO & $z_\text{em}$ & Instrument & SNR & Survey & Notes & Discovery ref. & Spectrum ref.\\
\hline
J1120+0641 & 7.0842 & X-Shooter & 35.0 & UKIDSS & & (1) & (26) \\
J1205-0000 & 6.8 & FOCAS & 3.5 & SHELLQs & & (20) & -- \\
J0224--4711 & 6.50 & GMOS & 6.5 & DES--VHS & &  (2)  & --\\
J0210--0456 & 6.44 & ESI & 5.3 & CFHQS & new reduction & (3) & (32)\\
J2329--0301 & 6.43 & ESI & 6.5 & CFHQS & new reduction & (11) & (25) \\
J1148+5251 & 6.419 & HIRES & 29.7 & SDSS & & (4) & (22) \\
J1152+0055 & 6.37 & FOCAS & 3.1 & SHELLQs & & (20) & --\\
J1148+0702 & 6.339 & HIRES & 3.4 & SDSS & & (5) & --\\ 
J0100+2802 & 6.30 & X-Shooter & 85.2 & SDSS & new spectrum & (6) & this paper\\ 
J1030+0524 & 6.28 & X-Shooter & 28.0 & SDSS & new reduction & (7) & (22)\\ 
J0050+3445 & 6.25 & ESI & 24.4 & CFHQS & & (3) & (23)\\ 
J0323--4701 & 6.25 & EFOSC & 12.5 & DES--VHS& & (2) & --\\
J0330--4025 & 6.25 & GMOS & 12.1 & DES--VHS& & (2) & --\\ 
J1623+3112 & 6.247 & ESI & 16.1 & SDSS &  & (8) & (22) \\
J2325-???? & 6.23 & EFOSC & 1.8 & DES--VHS &  & (24) & -- \\
J0410--4414 & 6.21 & GMOS & 12.7 & DES--VHS& & (2) & --\\
J0227--0605 & 6.20 & ESI & 7.5 & CHFQS & new reduction & (27) & (25) \\ 
J1048+4637 & 6.198 & HIRES & 29.2 & SDSS &  & (4) & (28)\\ 
J1609+3041 & 6.16 & MMT & 6.1 & SDSS & & (5) & --\\
J2229+1457 & 6.15 & ESI & 6.0 & CHFQS & new reduction & (3) & (25) \\ 
J1250+3130 & 6.13 & ESI & 26.2 & SDSS & new reduction & (9)\\
J0033--0125 & 6.13 & ESI & 6.1 & CHFQS & new reduction & (11) & (25) \\
J1319+0950 & 6.132 & X-Shooter & 96.8 & UKDISS/SDSS & & (10) & (23) \\ 
J1509--1749 & 6.12 & X-Shooter & 88.9 & CFHQS &  & (11) & (22) \\ 
J2315--0023 & 6.117 & ESI & 29.8 & SDSS &  & (12) & (23) \\ 
J0454--4448 & 6.10 & MagE & 5.8 & DES & & (21) & --\\
J0109--???? & 6.10 & EFOSC & 4.2 & DES--VHS& &(24) & --\\
J2216-0016 & 6.10 & FOCAS & 2.4 & SHELLQs & & (20) & -- \\
J1602+4228 & 6.09 & MMT & 33.3 & SDSS & new reduction & (8) & (25)\\ 
J0303--0019 & 6.078 & ESI & 8.0 & SDSS & new reduction & (12) & (32)\\
J0353+0104 & 6.072 & ESI & 80.7 & SDSS &  & (12) & (23) \\
J2054--0005 & 6.062 & ESI & 39.5 & SDSS &   & (12) & (23) \\ 
J1630+4012 & 6.058 & MMT & 17.0 & SDSS &  & (4) & (22)\\
J1641+3755 & 6.04 & ESI & 9.0 & CHFQS & new reduction & (11) & (25) \\
J0408--5632 & 6.03 & EFOSC & 4.3 & DES--VHS& & (2) & --\\
J1257+6349 & 6.02 & MMT & 6.1 & SDSS & & (13) & --\\
J1306+0356 & 6.016 & X-Shooter & 55.8 & SDSS & new reduction & (7) & (22)\\ 
J1137+3549 & 6.01 & ESI & 31.7 & SDSS & new spectrum & (9) & this paper\\
J2310+1855 & 6.003 & LBT-MODS & 17.9 & SDSS &  & (5) &-- \\
J0818+1722 & 6.0 & HIRES & 39.8 & SDSS &  & (9) & (28)\\
J0131--???? & 6.00 & EFOSC & 3.9 & DES--VHS& & (24) & --\\
\end{tabular}}
\label{masterfile}
\vspace{-0.5em}
\caption{Data used in this work. References are given in the caption of Fig~\ref{table:p2}. A dash `--' indicates the discovery spectrum is used. Question marks in quasar names indicate a quasar yet unpublished by the discovering authors.}
\end{table*}

\begin{table*}
\resizebox{\textwidth}{!}{%
\begin{tabular}{l c c c c c c c}
\hline
QSO & $z_\text{em}$ & Instrument & SNR & Survey & Notes & Discovery ref. & Spectrum ref.\\
\hline
J0841+2905 & 5.96 & ESI & 11.2 & SDSS &  & (15) & (22)\\ 
J0122-???? & 5.96 & EFOSC & 3.0 & DES--VHS& & (24) & --\\
J1202--0057 & 5.93 & FOCAS & 2.2 & SHELLQs & & (20) & --\\
J0008--0626 & 5.929 & MMT & 4.4 & SDSS & & (13) & -- \\
J1411+1217 & 5.927 & ESI & 15.9 & SDSS & & (8) & (22)\\
J0148+0600 & 5.923 & X-Shooter & 128.0 & SDSS &  & (13) & (23)\\ 
J1335+3533 & 5.901 & ESI & 16.3 & SDSS &  & (9) & (22)\\ 
J2119--0040 & 5.87 & MMT & 4.0 & SDSS & & (5) & --\\
J2307+0031 & 5.87 & MMT & 3.6 & SDSS & & (5) & -- \\
J0850+3246 & 5.867 & MMT & 3.7 & SDSS & & (13) & --\\
J0203+0012 & 5.86 & ESI & 17.4 & UKIDSS/SDSS &  & (12) & (23)\\
J0005--0006 & 5.850 & ESI & 28.8 & SDSS & new reduction & (8) & (32) \\ 
J1243+2529 & 5.85 & MMT & 4.1 & SDSS & & (5) & --\\
J0840+5624 & 5.844 & ESI & 17.6 & SDSS &  & (9) & (22)\\
J1436+5007 & 5.83 & MMT & 3.2 & SDSS & & (9) & --\\ 
J0239--0045 & 5.82 & MMT & 4.9 & SDSS & & (16) & --\\
J0836+0054  & 5.810 & X-Shooter & 93.4 & SDSS & new reduction & (7) & (22)\\ 
J0002+2550 & 5.8 & HIRES & 71.7 & SDSS & & (8) & (28) \\ 
J0810+5105 & 5.80 & MMT & 10.0 & SDSS & & (5) & -- \\
J1044--0125 & 5.782 & ESI & 49.2 & SDSS & new reduction & (17) & (25)\\
J0927+2001 & 5.772 & X-Shooter & 73.7 & SDSS & new spectrum & (9) & this paper\\
J1621+5155 & 5.71 & MMT & 10.3 & SDSS & & (5) & --\\
\end{tabular}}
\label{table:p2}
\caption[Table V.1, cont. ]{Current list of quasars, continued. Quasar names including question marks are not public yet. References: (1) \citealp{Mortlock11}; (2) \citealp{Reed17}; (3) \citealp{Willott10}; (4) \citealp{Fan03}; (5) \citealp{Jiang16}; (6) \citealp{Wu15}; (7) \citealp{Fan01}; (8) \citealp{Fan04}; (9) \citealp{Fan06}; (10) \citealp{Mortlock09}; (11) \citealp{Willott07}; (12) \citealp{Jiang08}; (13) \citealp{Jiang15}; (14) \citealp{Carnall15}; (15) \citealp{Goto06}; (16) \citealp{Jiang09}; (17) \citealp{Fan2000}; (18) \citealp{Wang16}; (19) \citealp{Morganson12}; (20) \citealp{Matsuoka16}; (21) \citealp{Reed15}; (22) \citealp{McGreer15}; (23) \citealp{Becker15}; (24) Reed in prep.; (25) \citealt{KOA}; (26) \citealt{Bosman17}; (27) \citealt{Willott09}; (28) \citealt{Becker06}; (29) \citealt{Venemans13}; (30) X-Shooter archives; (31) \citealt{Venemans15}; (32) \citealt{Eilers17}}
\end{table*}

\subsection{SDSS quasars}

The SDSS is a sky survey over 14,555 $\text{deg}^{2}$ which provides imagining in the \emph{ugriz} photometric bands as well as spectroscopic follow-up using a 2.5m dedicated telescope located at Apache Point Observatory \citep{SDSS, SDSS2}. Here we briefly outline the detection procedure of quasars in the SDSS (see \citealt{Jiang16} for a more in-depth summary). Candidates are selected in the first step as drop-outs with no detections in the \emph{ugr} photometric bands and with colours in excess of $i_{AB} - z_{AB} > 2.2$. After quality cuts, follow-up photometry is obtained in the near infra-red (IR) and a second cut $z_{AB} - J < 0.5 + 0.5\Delta_{i-z} $ is imposed (e.g. \citealt{Fan99}). Alternative colour cuts are used in deeper areas of the survey near the Galactic cap (\citealp{Jiang08, Jiang09}) and in regions scanned two or more times (\citealp{Jiang15}).

Confirmation spectra of the quasar candidates are typically obtained obtained with the Red Channel Spectrograph (RCS) on the 6.5m Multiple Mirror Telescope (MMT) or Double Spectrograph on the Hale 5.1m telescope (DBSP) (e.g. \citealp{Jiang16}), and in one occasion with the Multi-Object Double Spectrographs for the Large Binocular Telescope on Mt. Graham in south-eastern Arizona (LBT-MODS, \citealt{LBTMODS}). Additional near-IR spectra taken for some objects do not cover the range 7500\AA \ -- 10,000\AA \ required for coverage of \lal at $5.3<z<7.0$ and are not used in this work (e.g. \citealp{Jiang07}, \citealp{Simcoe11}).

\cite{Jiang16} presented the 52 final quasars discovered by the SDSS, most of which are included in this work. Out of those, 29 have been re-observed since their discovery to obtain higher quality data, while 23 have not. The discovery spectra for those 13 of these objects are included in our sample. Eight of those objects were reported in \citet{Jiang16}, three objects in \citet{Jiang15}, one objects in \citet{Jiang09} and one object in \citet{Fan06}. 
 
\subsection{DES and DES--VHS quasars}

The Dark Energy Survey (DES) covers an area of 5000 $\text{deg}^{2}$ in the southern hemisphere in visible imaging. It employs the dedicated Dark Energy Camera (DECam) on the Blanco 4m telescope, Cerro Tololo \citep{DES}. The first high-$z$ quasar discovered in DES was presented in \citet{Reed15}.
Quasar candidates are selected using a drop-out technique similar to the SDSS procedure described above, this time with the condition $i_{DES} - z_{DES} > 1.694$. In addition, the DES survey includes the \emph{Y} band, allowing a more efficient removal of red dwarves from the sample via a constraint on the quasar continuum of $z_{DES} - Y_{DES} < 0.5$. In \citet{Reed17}, eight additional quasars were detected by combining DES data with infrared observations in overlapping footprint of the VISTA Hemisphere Survey (VHS; \citealt{VHS}). Nine additional objects have been discovered in the same way since (Reed et al, in prep), of which four included here.

Spectroscopic confirmation of the candidates was conducted either with the ESO Faint Object Spectrograph and Camera (EFOSC, \citealt{EFOSC}) or the Gemini Multi-Object Spectrographs (GMOS, \citealt{GMOS}), with some objects subsequently observed in higher quality with the Magellan Echellette (MagE, \citealt{MagE}). The best quality spectrum for each of the 10 DES--VHS quasars was chosen as shown in Table~1.

\subsection{SHELLQs quasars}

The Subaru High-z Exploration of Low-luminosity Quasars (SHELLQs, \citealt{Matsuoka16}) is a new imaging survey utilising the Hyper Suprime-Cam on the Subaru 8.2m telescope \citep{Miyazaki12}. A search for quasars has currently been conducted over an area of 430 $\text{deg}^{2}$. The SHELLQs project aims to obtain deeper exposures in the \emph{grizy} bands compared to SDSS and DES, leading to the discovery of 33 faint $z>5.7$ quasars so far \citep{Matsuoka16, Matsuoka17}. In this work, we include four out of the first nine SHELLQs quasar spectra presented in \citet{Matsuoka16}. The confirmation spectra for these objects were taken with the Faint Object
Camera and Spectrograph on the Subaru telescope (FOCAS, \citealt{FOCAS}) as described in the discovery paper.

\subsection{Other quasar spectra}

In this work, we re-use 20 quasar spectra presented in previous investigations of \lal opacity. 
\citet{McGreer11} and \citet{McGreer15} conducted observations of 22 previously known quasars with the MagE, MMT and the X-Shooter instrument on Cassegrain UT2 \citep{Vernet11}. 
We are making use of 9 of those observations as indicated in Table 1. 
Similarly, \citet{Becker15} published spectra of seven quasars, at $5.98 < z < 6.25$, not included in a previous work by Fan et al. (2006), obtained on the Echellette Spectrograph and Imager (ESI) on the Keck II telescope \citep{ESI}, and X-Shooter on the VLT. Three additional spectra were first presented in \citet{Becker06}.
The quasars followed up in the above papers were initially discovered by various surveys including the UKIRT Infrared Deep Sky Survey (UKIDSS, \citealt{UKIDSS}), the Canada-France High-z Quasar Survey (CFHQS, \citealt{Willott07}), SDSS, and the Panoramic Survey Telescope and Rapid Response System (Pan-STARRS, \citealt{Morganson12}). In addition, we also include a 30h X-Shooter spectrum of \ulas at $z=7.08$, first presented in \citet{Bosman17}.

\subsection{New reductions}

After a proprietary period of 18 months, raw data obtained with ESI is made publicly available through the Keck Observatory Archive (KOA\footnote{\url{https://koa.ipac.caltech.edu/cgi-bin/KOA/nph-KOAlogin}}). In this work, we use 7 ESI spectra of quasars at $5.85<z<6.43$ re-reduced from raw data obtained from the KOA. Our ESI reduction pipeline is the same as described in Subsection~\ref{subsec:new}.


Finally, we re-reduced the spectra of five quasars which have been previously published. X-Shooter spectra of the quasars J0836+0054 and J1030+0524 were introduced in \citet{McGreer15}, but here we use our own reduction of the raw X-Shooter files instead, in an attempt to improve data quality. ESI spectra of quasars J0210-0459, J0303-0019, and J0005-0006 were part of an observing run (PID: C197E; PI: Sargent) in October 2010 and have been previously used in e.g. \citet{Eilers17}. Here we use our own reductions of the raw ESI data.

\begin{table*}
\begin{tabular}{c c c c c c c}
\hline
\hline
Object & $z$ & Instrument & Date & Exposure & Slit Width & Seeing \\
 & & & & Time (s) & (arcsec) & (arcsec) \\
\hline
SDSS J0100+2302 & 6.3 & X-Shooter & 23rd Oct 2015 & 1800 & 1.00 & 0.80\\
SDSS J1137+3549 & 6.01 & ESI & 18th Mar 2016 & 3000 & 1.00 & 0.80 \\
SDSS J0927+2001 & 5.772 & X-Shooter & 13th Jan 2010 & 1800 & 1.00  & 0.77\\
\end{tabular}
\label{table:1}
\caption{New observations of four $z>5.7$ quasars which are were not presented in previous work.}
\end{table*}

\subsection{New spectra}\label{subsec:new}

We present 3 new observations of quasars which were carried out on the ESI, X-Shooter and EFOSC instruments as detailed in Table 3. The spectra are plotted in Figure~1.

An X-Shooter spectrum was obtained in January 2010 of quasar SDSS J0927+2001 at $z=5.772$ and had not previously been published. The reduction procedure is the same as the one presented in e.g. \citet{Becker15}. The spectrum was extracted optimally \citep{Horne86} using 10 km $\text{s}^{-1}$ bins after being flat-fielded and sky-subtracted following \cite{Kelson03}. Custom telluric absorption routines were used as presented in \citet{Becker12}. 

We obtained a 3000s ESI spectrum of the $z=6.01$ quasar SDSS J1137+3549 on the 18th of March, 2016.


A deep X-Shooter spectrum of J0100+2302 was obtained in collaboration with Max Pettini over 2015 and 2016 as part of the 13 hours program 096.A-0095(A). Here we make use of only one exposure of 1800s of the object, which is nevertheless a great improvement upon the previous LBT-MODS spectrum of the quasar.

\begin{figure*}
\centering
\includegraphics[width=0.9\textwidth]{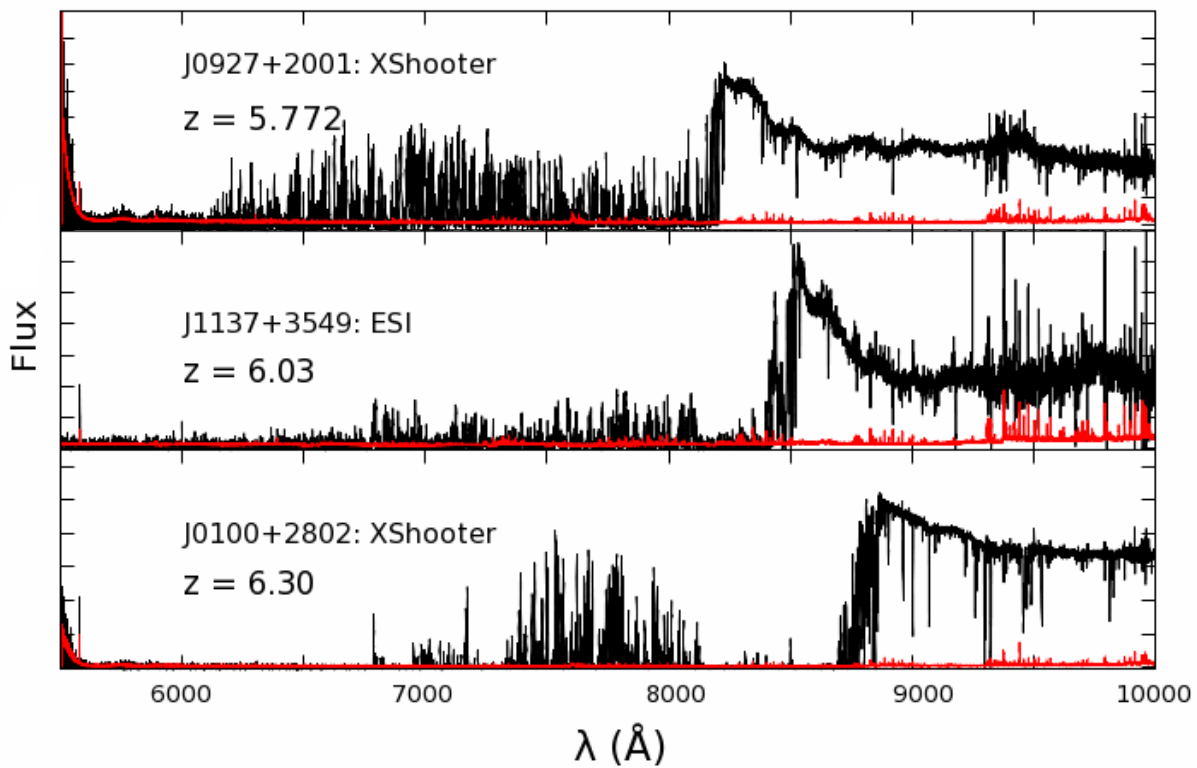}
\caption{New quasar spectra used in this work. Details of the observations can be found in Table 3.}
\end{figure*}

\subsection{Sample properties and notes on individual objects}

\begin{figure}
\centering
\includegraphics[width=0.8\columnwidth]{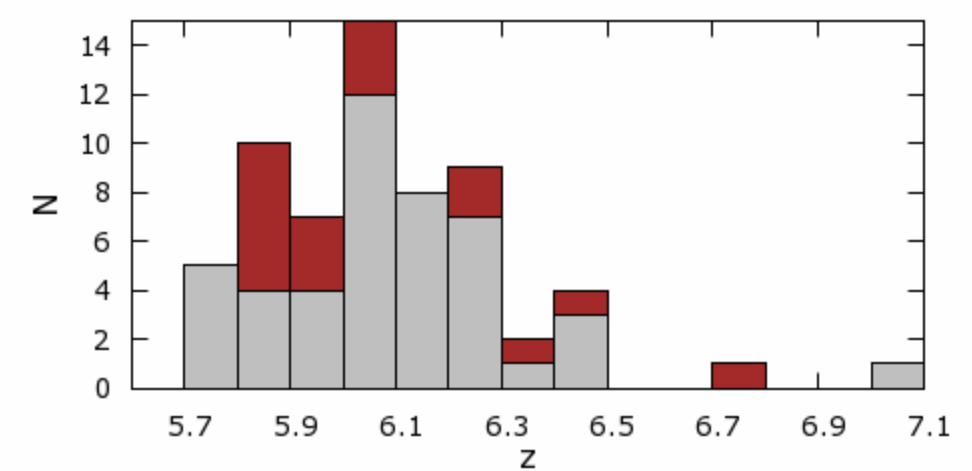}
\caption{Redshift distribution of the quasars included in our sample. Red indicates the whole sample while grey corresponds to the `SILVER' sub-sample of objects with SNR$>5$.}
\label{z_dist}
\end{figure}

\begin{figure}
\centering
\includegraphics[width=0.8\columnwidth]{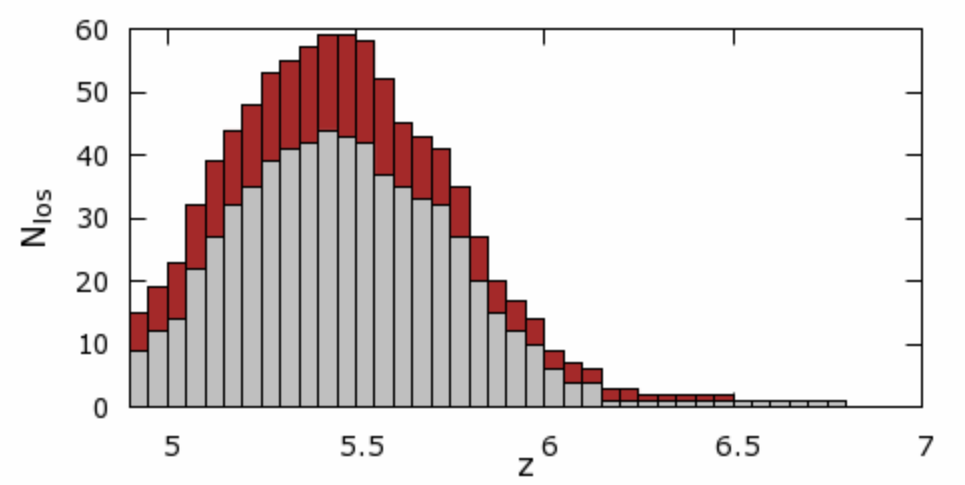}
\caption{Cumulative number of lines of sight covering a particular redshift. Red indicates the whole sample while grey corresponds to the `SILVER' sub-sample of objects with SNR$>5$.}
\label{cumm}
\end{figure}

Our 62 quasars have source redshifts in the range $5.70<z<7.08$ with a peak at $z\sim6.0$ and a distribution as shown in Fig~\ref{z_dist}. We investigate the \lal forest over  $1041<\lambda_\text{rest}<\sim 1178$\AA, resulting in a redshift coverage shown in Fig~\ref{cumm} with up to 59 
lines of sight covering the interval $5.4<z<5.5$. These distributions vary slightly depending on the choice of proximity-zone cut-off.


Redshifts for the objects in our sample are based the on best available estimates using nebular lines (such as [C{\small{II}}]) whenever possible. See \cite{Jiang16} for the origin of the measured redshifts of SDSS quasars. Since we are primarily concerned with measuring \lal \ opacity in the IGM, accurate values for the quasar redshifts are only relevant to the exclusion of quasar proximity zones from the analysis.
In this work we are not attempting to measure the evolution of quasar proximity zone length across redshift. We use a fix cut--off for the end of the proximity zone of $\lambda_\text{end, prox} = 1178$ \AA \ after checking that this is a reasonable choice and that more stringent criteria do not affect the results. This analysis is presented in Section~\ref{sec:prox}.

We list the signal-to-noise ratio (SNR) for our spectra in the fourth column of Table~1. The SNR measurement is complicated by the disparate resolutions and redshifts of the quasar spectra, and because common sky lines fall at different rest wavelengths across the sample. To measure SNR, we first normalise the spectrum by a power law as described below, then keep the pixels located at $1275<\lambda<1285$ \AA \ which are not affected by sky lines. This wavelength range covers a portion of quasar spectra minimally affected by broad emission lines. 
We present here the mean SNR per 60 km s${}^{-1}$. The SNR is then computed as,
\begin{equation}
\text{SNR} = \left< \frac{F}{\epsilon} \right> \cdot \sqrt{N_{60}}
\end{equation}
where $\epsilon$ is the error and $N_{60}$ the number of pixels per 60 km s${}^{-1}$ interval. This is computed using $N_{60} = 60.0 / \Delta v$, where $\Delta v$ is the pixel size in km s${}^{-1}$. For spectra where bins of fixed wavelength interval, $\Delta \lambda$, are used, rather than a fixed velocity interval, $\Delta v$ is measured at $\lambda = 8000$ \AA.
While this is not the only way of homogeneously measuring spectral SNR, it is sufficient for our purposes to discriminate between data quality regardless of resolution, and has the advantage of being invariant under re-binning of the spectra.
The values obtained range from SNR$=1.8$ for J2325-5229, a DES--VHS quasar with continuum emission barely above the detection threshold, to SNR$=96.8$ in a deep X-Shooter exposure of J1319+0959 first used in \citet{Becker15}.

Out of the quasars in our sample, none show the characteristic features of broad absorption line spectra (BAL) or contamination by a DLA. Such objects were explicitely excluded during the sample assembly, with BAL and DLA features accounting for 5 out of 30 object rejections. 
We note that the fraction of quasar spectra displaying these features (5 out of 92 or about 5\%) is lower than measured at later times. This could be due to the selection techniques employed to discover high redshift quasars, as the presence of a BAL feature diminishes the photometric colours most commonly used to select quasar candidates as drop-outs. It is also likely that some contamination by DLAs has gone undetected in our sample. Due to the saturation of the \lal forest, the most reliable way to detect and remove DLAs from the sample would be by detecting associated metal absorption at the DLA redshift. \magtwo is unfortunately not visible for the redshifts of interest here, and the quality of the spectra is insufficient to detect typical \cfour absorption systems in the majority of cases. Only deep X-Shooter spectra would provide sufficient coverage and sufficient SNR to completely remove DLA contamination at $z>5.5$. 




\section{Methods}

The spectra are first normalised by fitting a power law to the unabsorbed continuum. The portion being fitted extends from 1270 \AA --1450 \AA \ in the rest frame of the quasar; the range 1270 \AA --1350 \AA \ is used instead when the spectral coverage stops short of 1500 \AA \ to avoid portions of the spectrum affected by the falling response of the instrument. This is the case for instance for spectra of $z>5.7$ quasars taken with the MMT or MagE, whose coverage extends to $\lambda=10,000$\AA \ but the response of which decays significantly from $\lambda\gtrsim9700$\AA. Pixels affected by sky lines are excluded and a first power law (PL) fit is made, from which we then exclude any pixels for which $|F_\text{QSO} - F_\text{PL}| > 2\epsilon_\text{QSO}$. The remaining flux is then fit with a power law again, and the process repeated a second time with a deviation coefficient of 1.5 to ensure convergence. Finally, the full flux array is divided pixel-by-pixel by the best fit power law function thus obtained.

\begin{figure*}
\centering
\includegraphics[width=\textwidth]{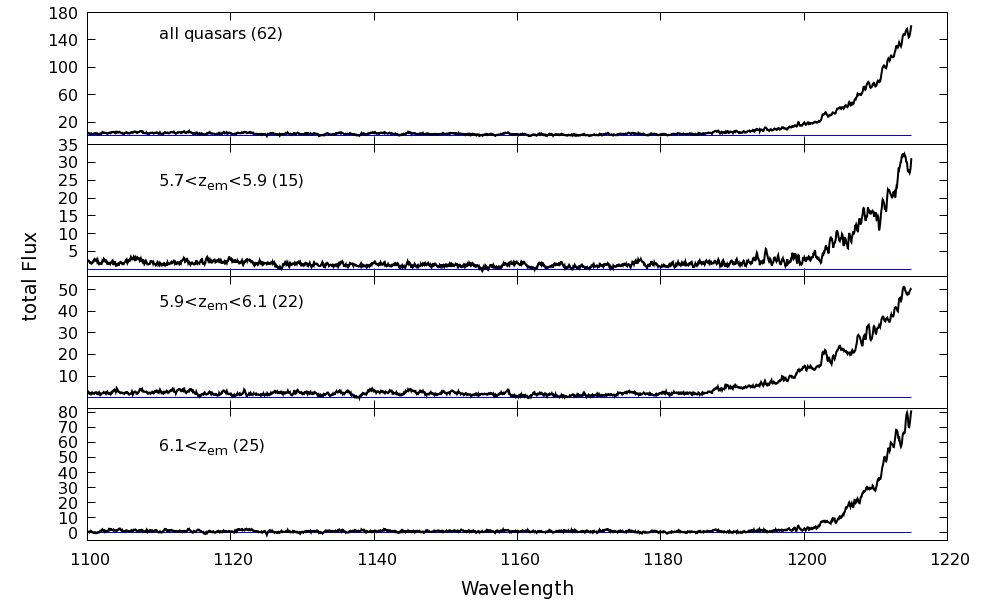}
\caption{Stack of flux over $\lambda=$1100 \AA --1215 \AA \ of 62 $z>5.7$ quasars. Spectra were normalised and sky lines masked before stacking. While the proximity zone transmission has fallen considerably by 1180\AA, the exact end of the host quasar's influence on the flux is unclear. Five spectra are excluded as described in Section V.2.7. Panels show the evolution of the stacked spectrum over $5.7<z<5.9$, $5.9<z<6.1$ and $6.1<z$. The thin blue lines show a total flux of zero.}
\label{stack}
\end{figure*}

Three objects which displayed too little continuum (due to a combination of high redshift and spectrograph wavelength coverage) are excluded from the analysis since no satisfactory estimate of the continuum could be obtained. The continua for all other objects were checked visually. We checked that the best fit power law parameters were robust to small changes in the fitting window bounds. The effects of window choices were run all the way through the analysis; we find an end effect on values of $\tau$ of magnitude $\Delta \tau\lesssim10\%$. This effect can be seen in Fig.\ref{compar}, where the only differences between our measurements and those of \citet{Becker15} are due to small differences in the choices of continuum fitting. These errors are in all cases much smaller than the effect of cosmic variance.

We measure the average transmitted flux in windows of 50 comoving cMpc $\text{h}^{-1}$ extending from the end of the quasar's proximity zone  at $\lambda_\text{rest} = 1178$\AA \ down to the onset of \lab \  absorption (1041 \AA \ in the rest frame). The average continuum-normalised flux is transformed into effective opacity following \red{ $\tau_\text{eff} = - \text{log}(\langle F\rangle)$} and associated to the redshift corresponding to the middle of that 50 cMpc $\text{h}^{-1}$ region. The analysis is repeated for window sizes of 10, 30 and 70 cMpc $\text{h}^{-1}$.

We treat non-detections of transmitted flux in two different ways. First, following previous work, 
we take the upper limit on the flux to correspond to twice the error in the flux over the measurement window. If individual peaks of transmission are detected at more than $2\sigma$ significance over that range, then we take the lower limit on the flux to be equal to twice the flux in those peaks alone following $F > F_\text{peaks} - 2\sigma_\text{peaks}$ where $\sigma_\text{peaks}$ is the error over the wavelength covered by peaks \citep{Becker15}. 
This allows us to compare our results to the previous samples of \cite{Fan06} and \cite{Becker15} which are a sub-set of our catalogue. 
The \lal opacities in these two samples were not measured in identical ways, as the lengths of the excluded proximity zones and the details of the continuum fitting were subtly different. This might have resulted in a mild tension between the two samples, which we can now harmonize by \begin{enumerate} \item doing a bootstrap re-sampling of our catalog to mimic previously used sample sizes, and \item treating both samples identically through the same continuum fitting routine and same proximity zone exclusion technique. \end{enumerate}

Secondly, we treat upper limits by plotting the most optimistic and pessimistic bounds on the cumulative distribution function (CDF). The optimistic bound is given, as above, by taking the intrinsic average flux to be equal to two times the average error over the measurement window, i.e. just below detection sensitivity. However, different spectrographs with different exposure times will be sensitive to different thresholds: for instance, none of the data in our sample could measure an effective optical depth \red{$\tau_\text{eff} \geq 8.0$}. To reflect this ambiguity, we attribute maximal opaqueness (\red{$\tau_\text{eff} \to \infty$}) to all non-detections in order to obtain a `maximally pessimistic bound' (so called due to the increased difficulty of reconciling this outcome with current reionisation models). The `true' CDF will necessarily lie in-between these two extremes so long as the bounds are resolved.

\begin{figure*}
\centering
\includegraphics[width=0.8\textwidth]{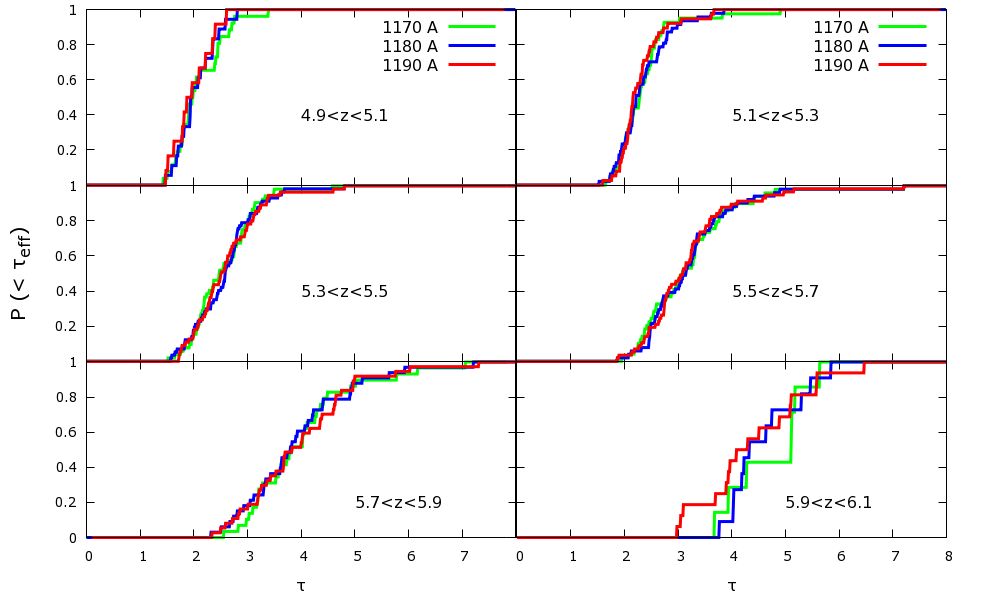}
\caption{Effect of incrementally increasing the excluded proximity zone size on the \lal CDF at various redshifts (see also Fig~4). By ending the proximity zone at  $\lambda = 1190$ \AA, the average opacity is affected at $z<5.3$ and $5.5<z<5.7$. However, no statistically significant difference is seen between a cut-off at  $\lambda = 1180$ \AA \ and the very conservative case  $\lambda = 1170$ \AA, at any redshift. Throughout this paper we adopt the traditional cut-off $\lambda = 1178$ \AA.} 
\label{prox}
\end{figure*}

\subsection{Proximity zone exclusion}\label{sec:prox}

We aim to measure the \lal opacity of the IGM. In order to achieve this, we need to avoid any bias introduced by the quasar proximity zones: the regions immediately surrounding the quasars where the ionisation of the gas is enhanced due to ionising radiation from  the object themselves. In the past, cut-offs for quasar proximity zones were chosen either on a case-by-case basis as the point where the quasar's \lal flux had fallen below 10\% of its peak value \citep{Fan06, Eilers17} or chosen as a fixed value of 1176 \AA \ determined based on a small sample of objects \citep{Becker15}. The former definition is not useful in the context of a sample containing spectroscopic data of varied resolution since clumps of neutral gas within the proximity zone might be resolved by some instruments but not others. A fixed-value cut-off is therefore more suited to a large dataset and facilitates future refinements of the measurements. 

To ensure that the traditional value of 1176\AA \ is sufficiently stringent as to remove all proximity zone influence, we first plot a stacked spectrum of our entire catalog in the range $1100< \lambda < 1215$ \AA \ (Figure~4). 
The stacking was carried out by interpolating the spectra onto
a common wavelength array after normalising them by a fitted power law and removing bad
pixels. For the purposes of the stacking, objects with different measurements errors are not
weighed differently.
Some interesting features are visible, such as the slight increase in average opacity along lines of sight with redshift and the average power-law shape of the proximity zone. Based on the stack, we see that the proximity zone influences average transmitted flux at $\lambda > 1190$\AA, but effects further away are unclear. To confirm and refine this, we incrementally increase the amount of excluded flux blueward of the \lal line and compare the resulting CDFs. This is equivalent to restricting the measurement to spectrum segments which are increasingly distant from the redshift of interest.
The results are shown in Fig~\ref{prox} for several redshift bins. The effect of the proximity zone flux is visible as a modest decrease in opacity when a cut-off of $\lambda = 1190$ \AA \ is used. 
\red{A two-sample KS test \citep{KS} yields levels of agreement $p>0.8$ between the $\lambda = 1180$ \AA \ and the  $\lambda = 1170$ \AA \ cases, indicating that the shape of the CDF has converged. Mild disagreement is obtained with the $\lambda = 1190$ \AA \ case at $z=5.8$ and $z=6.0$ ($p>0.50$ and $p>0.20$ respectively).}
We therefore adopt a value of the proximity zone end at  $\lambda = 1176$ \AA \ in the rest of the paper.

Individual quasars showing anomalously long proximity zones (whether due to extreme bolometric luminosities, location in an under--dense IGM region, or chance) could be present in the sample and would not appear in the stacks in Fig~4. The resulting contamination could potentially bias the average opacities to be too low. However, we do not find any such objects in the sample by visual inspection. Uncovering trends in quasar proximity zones is beyond the scope of this work, and it is enough for our purposes to confirm that no boost to average flux is seen at $\lambda<1176$ \AA.

\begin{figure*}
\includegraphics[width=0.9\textwidth]{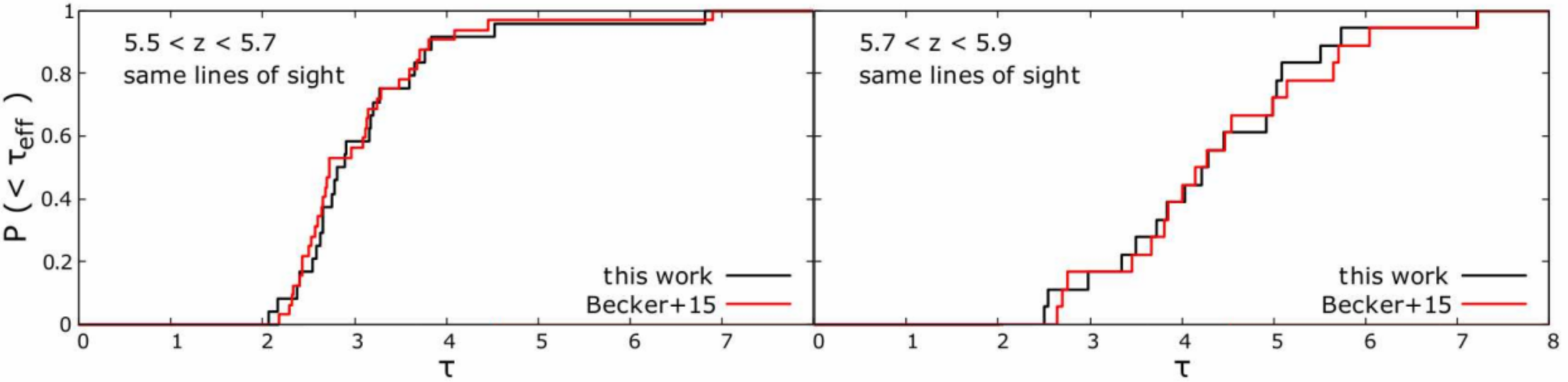}
\caption{Reproduction of the \citet{Becker15} opacity PDFs at $5.5<z<5.7$ (\emph{left}) and $5.7<z<5.9$ (\emph{right}). The same lines of sight are used, except for 4 quasars at $z<5.7$ that are excluded from our work for the lower redshift case. Differences are accounted for by slight differences in the continuum fitting between the authors and previous measurements taken from \citet{Fan06}.}
\label{compar}
\end{figure*}

\section{Results}

\begin{figure*}
\centering
\includegraphics[width=0.8\textwidth]{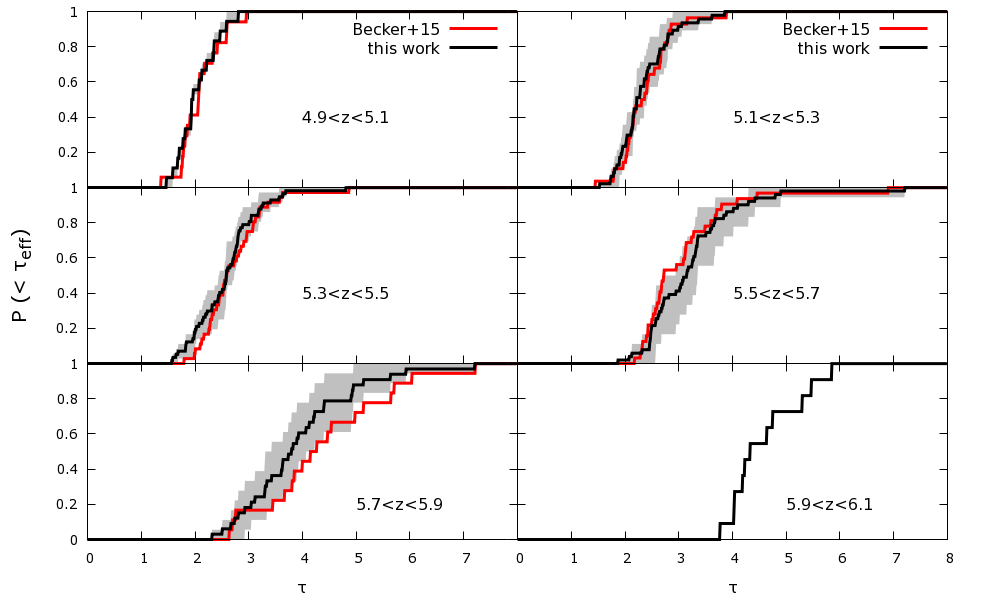}
\caption{New results, obtained using the method described in \citet{Becker15}. Only spectra of moderate or good quality (SNR$>5.0$, `SILVER' sample) are used, and non-detections of transmitted flux are treated as data points with values of twice the average error (see text). No significant discrepancy with previous results is found.}
\label{A}
\end{figure*}

\begin{figure*}
\centering
\includegraphics[width=0.8\textwidth]{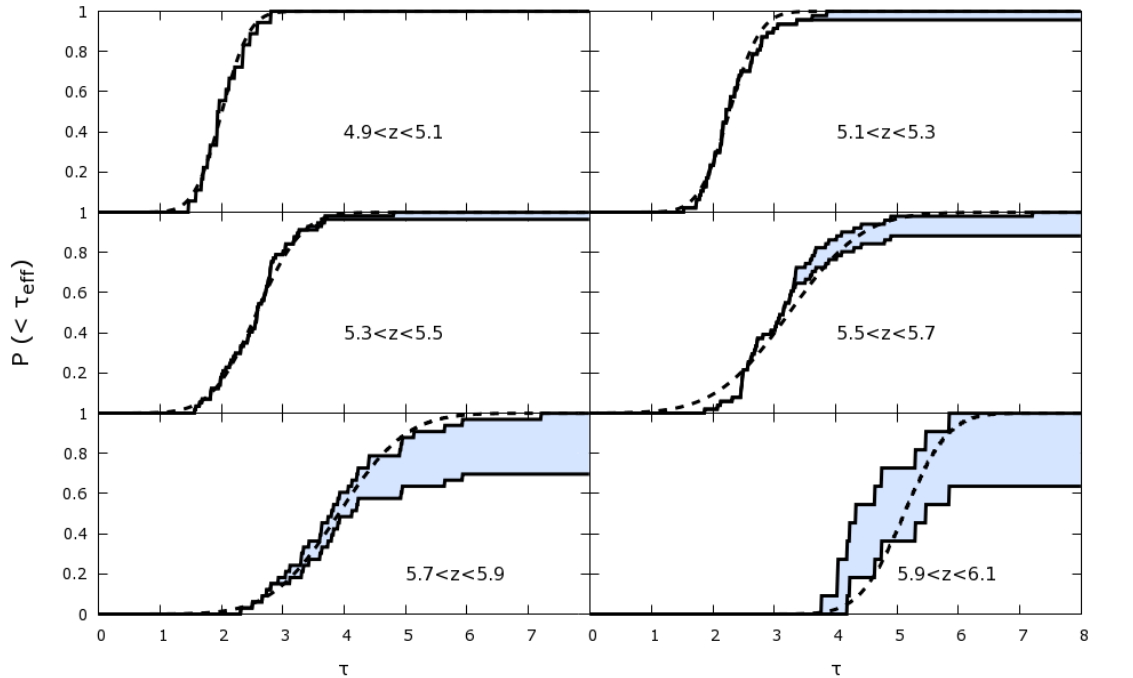}
\caption{New results, plotting the most optimistic and most pessimistic contours based on the intrinsic values values of non-detections. The leftmost contour corresponds to non-detections have intrinsic values of twice the average error (as in Fig~\ref{A}) while the rightmost contour assumes non-detections are maximally opaque (see text). The thin dashed line displays the most likely lognormal distribution computed in a maximum likelihood scheme (see Section 6).}
\label{B}
\end{figure*}

\subsection{Effect of data quality}

We select sub-samples from the quasar catalog based on the signal to noise ratios. A `GOLD' sample is chosen with SNR$\geq11$ to match the SNR of the worst spectrum used in \citet{McGreer15}. This yields a sample of 33 high-quality spectra, nearly halving the sample. 
Similarly, we construct a `SILVER' sample by applying a less stringent cut of SNR$\geq5$, matching the data quality from \citet{Eilers17}. This yields 45 lines of sight. For consistency, we will refer to the full sample as the `BRONZE' sample.
We note that the sample of 26 quasars used by \citet{Becker15}, the largest one so far, includes measurements from \citet{Fan06} made from spectra with SNR much lower than these thresholds -- down to SNR$=3.2$.
The results are shown in Figure~\ref{SNR}. There is no significant trend with data quality at $z<5.7$, where the distribution is well-sampled even by the small GOLD sample \red{(two-sample KS test $p>0.8$)}. A disparity appears at $z>5.7$. 

The differences between the SILVER and GOLD samples at $5.7<z<5.9$ are due to the small sample size of the GOLD sample. The difference in number of lines of sight in this range is a factor of 2.5, which reflects the fact that very high quality spectra are harder to obtain for quasars at $z>6.0$. The SILVER and GOLD sample are marginally consistent according to a KS-test ($p<0.2$). 

On the other hand, the sizes of the SILVER and GOLD samples are similar at $5.9<z<6.1$, and agree with each other well. However, the additional lines of sight included in the BRONZE sample are of insufficient quality to distinguish opacity beyond $\tau_\text{eff}\gtrsim3$. Since the mean opacity measured in the SILVER sample is roughly $\tau_\text{eff}\sim4.5$, these SNR$<5$ spectra yield upper limits (equal to twice the error) which are very poor in comparison. \red{The small sample size in this bin (8 measurements in the GOLD sample, 16 in the BRONZE one) mitigates the significance of this discrepancy, with the KS level indicating only mild discrepancy ($p<0.2$). In addition, two} of the BRONZE lines of sight do show signs of residual transmission (see below).

In light of these results, we decide to adopt the SILVER distribution in every instance in this paper where the data is presented or analysed in the form of a cumulative distribution function. The maximum likelihood method described above has been explicitly designed to account for observational errors and sensitivity, and we accordingly use the BRONZE sample only in Section 6.



\subsection{New \lal distributions}

Figure~\ref{A} presents our results compared to the previous CDFs of \citet{Becker15}. As done in \citet{Becker15}, the CDFs include lower limits. Figure~\ref{B} presents the same results, plotted to show the `pessimistic' and `optimistic' bounds described earlier. The results over $4.9<z<5.7$ are completely consistent with previous studies. We find a clear, well-defined tail of high-opacity ($\tau_\text{eff}>3$) lines of sight at redshift $z\sim5.2$. This trend was already visible in the CDF reported in \citet{Becker15}. 
Roughly $\sim20\%$ of lines of sight at $z=5.2$ have opacities $\tau_\text{eff}>2.5$, which might pose problems for IGM models that assume a spatially uniform UV background and temperature-density relation.

At $z\sim5.6$, we find a small but significant tail of transparent lines of sight, with roughly $\sim20\%$ of measurements showing $\tau_\text{eff}<2.5$. This tail was not visible in the \citet{Becker15} sample as most of the relevant objects were not included. \red{The two samples are consistent according to a two-sample KS test at all redshifts ($p>0.80$)}. At $z\sim5.8$, we find that opacities are slightly smaller than the ones previously reported. Our sample for $z\sim5.8$ is of comparable size to \citet{Becker15}'s samples at $z\sim 5.2$ and $z\sim5.4$, so that small differences are expected between our measurement and a `true' representation of cosmic variance in the same way as seen at lower redshifts. 

At $z\sim6.0$, our sample is smaller than all the ones used in \citet{Becker15} at lower redshift and the results should be interpreted with caution. Seven out of eleven 50 cMpc regions included in our SILVER sample display residual peaks of transmission, while the remaining four pose tight constraint on transmission. We find a very high average opacity of $\tau_\text{eff} \sim 4.5$. Such opacities are only accessible to current spectrographs with large time investments. This is readily visible in the large panel of Figure~\ref{SNR}, where the `transparent' lines of sight in the BRONZE sample are upper limits originating from spectrographs which struggle to distinguish opacities beyond $\tau \sim 3$. It is worthy to note that two lines of sight with SNR$<5$ do display residual transmission at the level of $\tau_\text{eff}\sim2.5$. The $z=6.34$ quasar J1148+0702 displays a transmission peak within $6$\AA \ of the formal end of the proximity zone, while the $z=6.23$ quasar J2325 displays such a peak outside of its proximity zone but has a SNR of 1.8, making it impossible to definitely rule of reduction issues. Further scrutiny of these objects is required in order to determine whether these peaks may be related to particularly long and sporadic proximity zones.



\begin{figure*}
\centering
\includegraphics[width=0.8\textwidth]{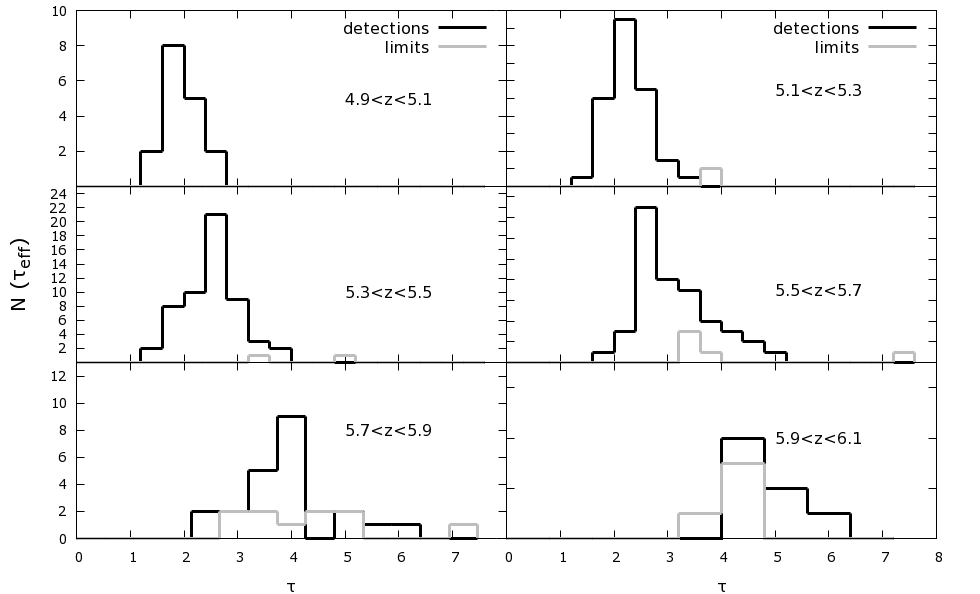}
\caption{Differential distributions of opacity in 50 cMpc $\text{h}^{-1}$ bins, showing detections in black and lower limits in gray.}
\label{PDF}
\end{figure*}

In Figure~\ref{PDF} we plot the distribution function of \lal opacity across redshift. We distinguish between detections and lower limits using separate histogram colours. The distributions are clearly non-gaussian, with peak values increasing linearly with redshift. The tail of opaque lines of sight at $z\sim5.2$ is clearly visible and appears smooth and well sampled.

\begin{figure*}
\centering
\includegraphics[width=0.8\textwidth]{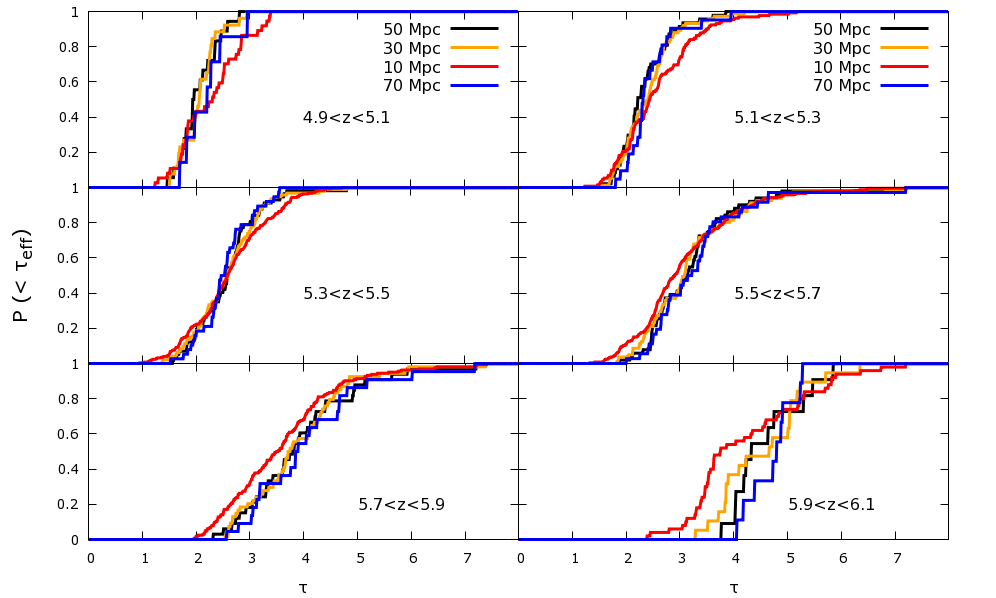}
\caption{The effect of varying the size of the window over which \lal transmission is measured. No significant effect is seen between windows of 30, 50 and 70 cMpc $\text{h}^{-1}$ at any redshift, suggesting that fluctuations occur on even larger scales. Binning the data in 10 cMpc $\text{h}^{-1}$ windows strongly affects the distribution, in particular at lower redshifts where it results a broader distribution.}
\label{C}
\end{figure*}

\begin{figure*}
\centering
\includegraphics[width=0.8\textwidth]{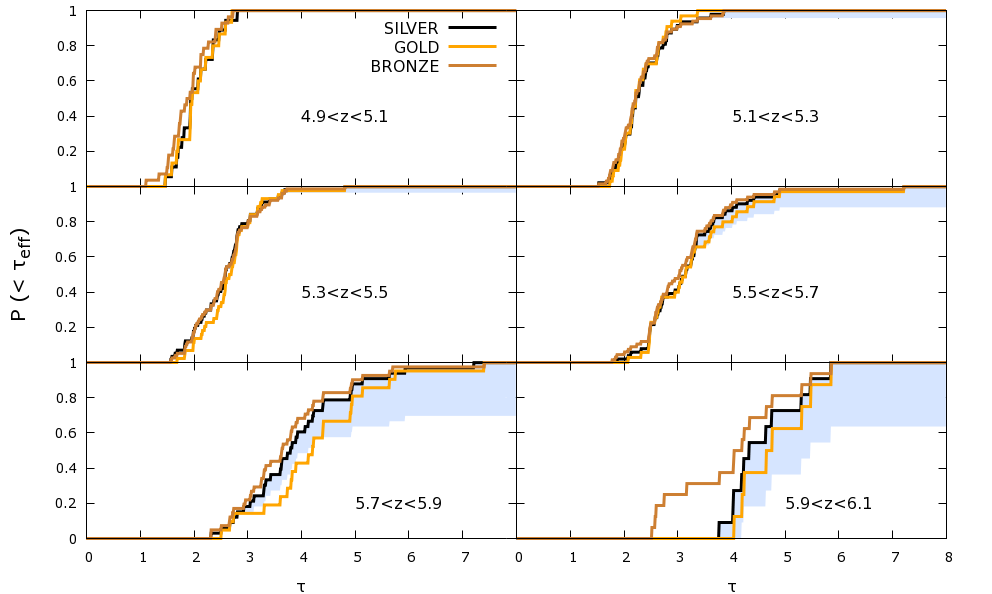}
\caption{Cumulative distribution functions of \lal opacity across redshift, computed using the full sample (black), the SILVER sub-sample of 51 objects matching the quality of data used in \citet{Eilers17} (SNR$\geq5.3$), and the GOLD sample of 35 objects which match the quality from \citet{McGreer15} (SNR$\geq11.2$). Shaded blue areas show the `optimistic' and `pessimistic' bounds presented in Fig.~\ref{B}. At $z<5.5$, the distributions are well resolved by all samples and the distributions therefore agree. At $5.5<z<5.9$, the GOLD distribution lies between the bounds shown in Figure~\ref{B}. At $z>5.9$, the difference is attributable to the small sample size of the sub-samples (see text).}
\label{SNR}
\end{figure*}

The effect of varying the size of the integration window is shown in Figure~\ref{C}. Although the effect is subtle, decreasing the window size tends to broaden the distribution, as expected. This is a natural consequence of cosmic variance. The broadening is particularly pronounced when the window size is decreased below 30 cMpc $\text{h}^{-1}$. The redshift range $5.1<z<5.7$ is more clearly affected, possibly because these redshifts are better sampled. \red{We find no statistically significant difference between bins of 30, 50 and 70 cMpc $\text{h}^{-1}$ at any redshift ($p>0.8$). The distributions with binning sizes of 10 cMpc $\text{h}^{-1}$ are strongly discrepant, with a two-sample KS test finding a probability $p<0.01$ that they were drawn from the same sample as the distributions with larger binning windows.  }

\subsection{Comparison to previous studies}

As a first test of our procedure, we reproduce the CDF presented in \cite{Becker15} using the spectra of 24 out of 27 $z>5.7$ quasars used in that work, which are a subset of our catalog. Figure~\ref{compar} presents the results at $5.5<z<5.7$ and $5.7<z<5.9$. The latter measurement makes exclusive use of those 26 quasars, and therefore any deviations are due entirely to subtle differences in the continuum fitting -- such as the precise wavelength ranges used and the number of sigma-clipping iterations. In addition, the opacities of the spectra used in \cite{Fan06} were not recomputed in \cite{Becker15}, giving rise to a separate set of slight discrepancies. All of our measurements of $\tau_\text{eff}$ in 50 cMpc $\text{h}^{-1}$ windows agree within error with those quoted in the original papers. The \citet{Becker15} results at $5.5<z<5.7$ make use of 4
quasars at $z<5.7$ which are not used in our sample. Nevertheless the shapes of the cumulative PDFs are in excellent agreement, and the measurements of $\tau_\text{eff}$ in individual windows all agree within error.

Next, we compare our results with those \cite{Becker15} by computing the CDF obtained from a random sub-sample of the size used by those authors from our larger sample. The contours of one hundred such realisations are plotted in Figure~\ref{A}. 
At all redshifts we find the results are in statistical agreement, with a mild tension ($p<0.5$) at $5.7<z<5.9$ where our work finds a slightly lower average opacity.


\begin{table*}
\centering
\begin{tabular}{r c c c c}
$z$&$\langle F \rangle_\text{obs}$&\red{\citet{Chardin17}}&\citet{Keating17}& Sherwood \\
\hline
5.0& $0.135 \pm 0.012$ & N/A & 1.703 & 0.662\\
5.2& $0.114 \pm 0.006$ &N/A & 1.522 & 0.590\\
5.4& $0.084 \pm 0.005 $ &0.878 & 1.509 & 0.565\\
5.6& $0.050 \pm 0.005 $&0.848 & 1.135 & 0.590\\
5.8& $0.023 \pm 0.004 $&0.841 & 1.055 & 0.666\\
6.0& $0.0072 \pm 0.0018$&N/A & 0.944 & 0.796\\
\end{tabular}
\caption{Emissivity rescaling factors ($\alpha$) used to tune the simulations discussed here. The factors are chosen to match the observed flux $F$ following $\langle F \rangle_\text{obs} = \langle \text{exp}(-\alpha \cdot \tau_\text{los}) \rangle$ (see text). The first column gives the measured values of $\langle F \rangle_\text{obs}$. To obtain the most accurate measurement possible, only the GOLD sample was used. The errors are estimated using bootstrap re-sampling.}
\label{alpha}
\end{table*}

\section{Comparison with models}

In this section we compare the updated CDFs of \lal transmission in 50 cMpc $\text{h}^{-1}$ windows with three simulations. Here, we briefly describe these simulations and outline their most relevant features.

When comparing predictions from these numerical simulations to observational results, it is important to keep in mind a few caveats. Firstly, all of the following numerical models explicitly re-scale the ionising background intensities to match the observed mean effective optical depth of the observations presented in this paper, $\langle F \rangle_\text{obs}$, by freely choosing a parameter $A$ such that:
\begin{equation}
\langle e^{ - A \tau_i } \rangle = \langle F \rangle_\text{obs}. 
\end{equation}

At low redshift ($z\lesssim5.7$), this rescaling is small and is somewhat justified by the difficulty of self-consistently generating the ionising UVB. In the following analysis, all simulation snapshots have been rescaled in this way to observed $\langle F \rangle_\text{obs}$ for each redshift bin. The values of $\langle F \rangle_\text{obs}$ are computed using the SILVER sample of objects with SNR$>5$, and are shown in Table~4.

These simulations all use the UVB prescription of \citet{HM12} (HM12) as a starting point, which incorporates the best available estimates of the nature of ionising sources, the ratio of galaxy to quasar contributions, the escape fraction, the spectral shape of sources, and other factors. The precise values of these parameters are not known, and vary across redshift. The HM12 emissivity is therefore used as a best guess, and the final re-scaling of the emissivity reflects the known inaccuracy in the model. The rescaling is therefore simulation-- and redshift--dependent, but also resolution--dependent since the small-scale recombination and self-shielding effects necessary to calculate the UVB self-consistently are currently beyond the reach of numerical simulations. In Table \ref{alpha} we list the values used to rescale in optical depths in each case ($\alpha$), chosen to match the observed flux $F$ following $F = \langle \text{exp}(-\alpha \cdot \tau_\text{los}) \rangle$. At high redshift ($z\gtrsim5.9$) this rescaling procedure becomes less valid because the correction is not small. For instance, the mean flux values of the Sherwood simulation used here had to be re-scaled by a factor $>10$ at $z=6$. This reflects the fact that the HM12 background is a bad representation of the ionising emissivity at high redshift, and ideally should not be used. 

Secondly, the timing of reionisation is a free parameter in most of the following models. The Sherwood simulation, and the HM12 ionising background, were designed to fit \lal transmission at lower redshifts than explored here. Because the time evolution of \lal transmission is slower at $z<4.9$, the successful predictions of these cosmological simulations are usually robust to shifts up to $\Delta z_\text{Re} \sim 0.2$ over the redshift range considered here (see e.g. \citealt{Chardin17-spikes}). 

\subsection{Full forward modelling}

To meaningfully confront simulated \lal lines of sight with observations, it is important to post-process them in a way which mimics data. We take a full forward modelling approach when comparing the data to simulations, transforming simulated lines of sight into realistic observations before treating them in the same way as the data. This simplifies the comparison, and enables us to estimate the errors in an empirical way.

We implement this by selecting the same number of simulated lines of sight as present in the data at each redshift, and post-processing them with observational profiles. An observational profile consists of an instrumental resolution and an error array. Each selected simulated line of sight is randomly matched to such a profile drawn randomly from the observations relating to the redshift under study. The simulated spectrum is first mapped onto a corresponding wavelength array, convolved with a gaussian with width of the instrumental resolution and re-binned onto the same wavelength array as the observation. Finally noise is added randomly at each pixel following a gaussian distribution with width corresponding to the observed error in that pixel.

\begin{figure*}
\centering
\includegraphics[width=0.8\textwidth]{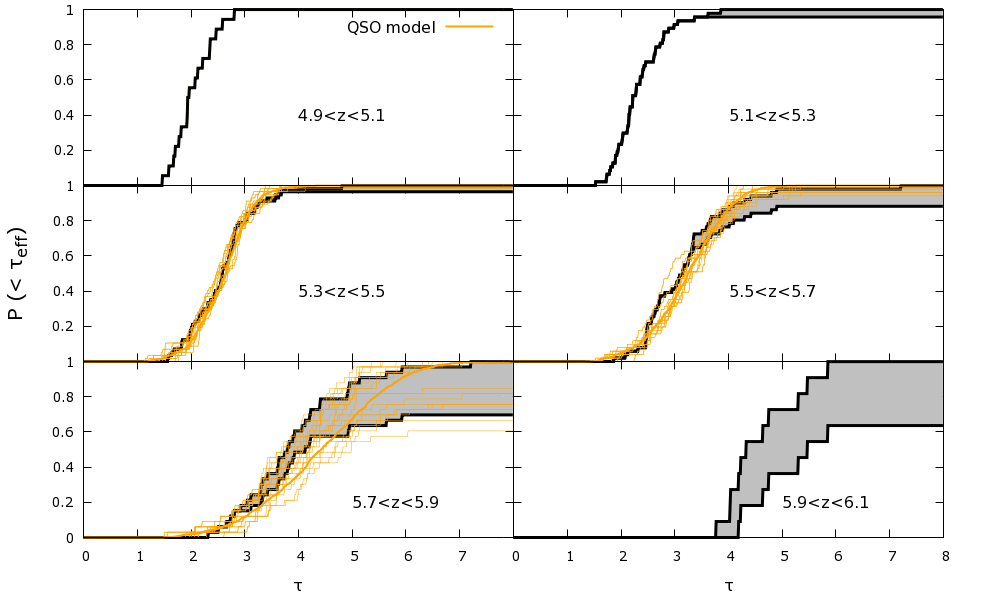}
\includegraphics[width=0.8\textwidth]{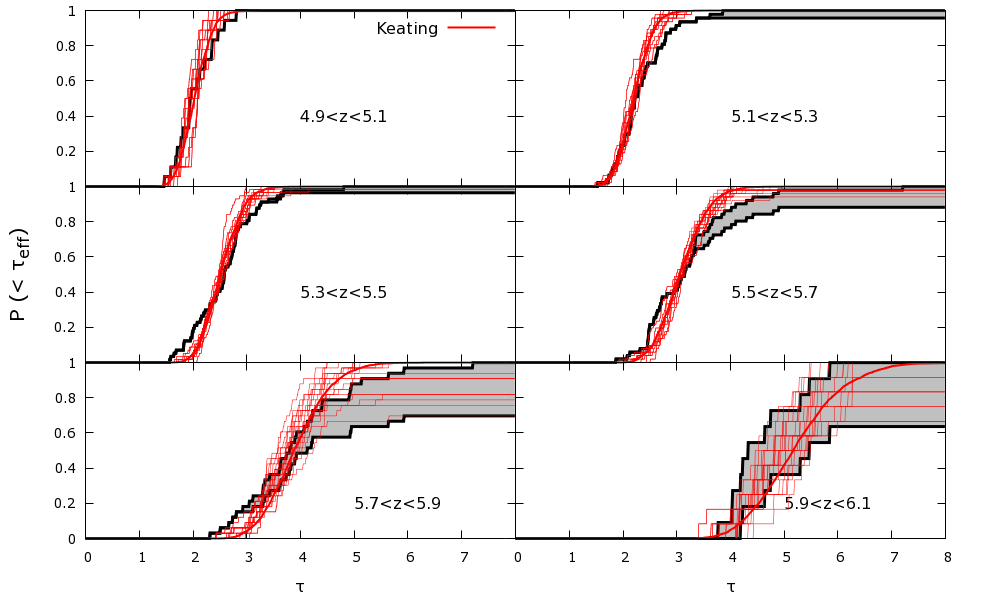}
\caption{Comparison of the measured \lal PDFs at $4.9<z<6.1$ with fully post-processed outputs of numerical simulations from \red{\citet{Chardin17}} (\textit{top}) and \citet{Keating17} (\textit{bottom}). The coloured contours show the envelope of the pessimistic and optimistic bounds for 100 realisations. Post-processing consists of randomly drawing a number of lines of sight from the simulations equal to the number of observations in the corresponding redshift range. The simulated lines of sight are then forward-modelled to mimic the observed spectra (see text).}
\label{sims_pp}
\end{figure*}

\begin{figure*}
\centering
\includegraphics[width=0.8\textwidth]{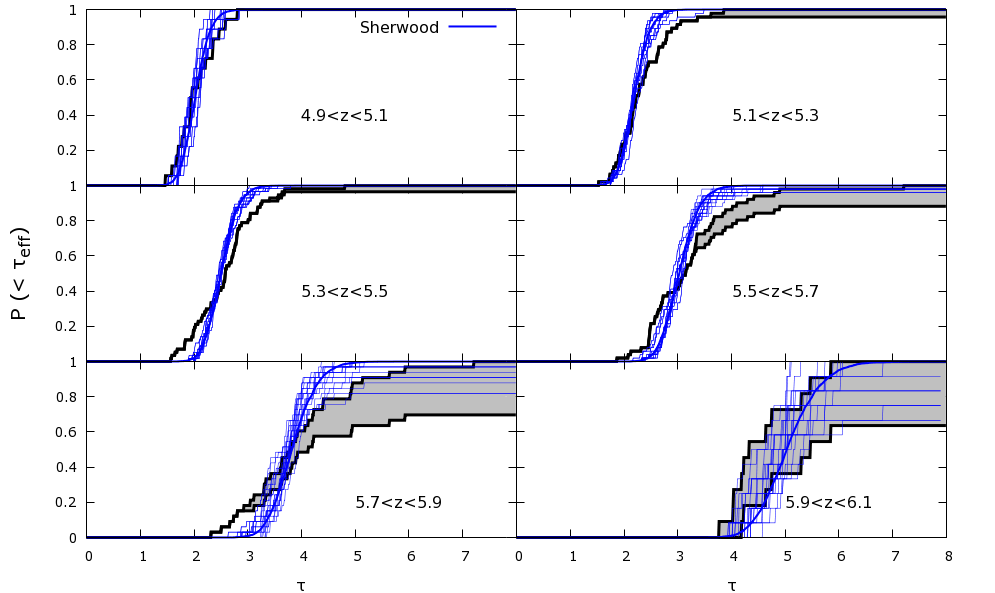}
\caption{Same as Figure~\ref{sims_pp} for lines of sight drawn from the Sherwood simulation \citep{Bolton17}.}
\label{sims_pp_2}
\end{figure*}

The resulting post-processed lines of sight are equal in number to the observations for the corresponding redshift range, and the CDF extracted from both sets of lines of sight should now in principle be completely comparable. To get an estimate of the variance between random realisations, we run the entire process 100 times until the envelope of the CDF bounds converges. CDFs outside of those bounds -- even by a single bin -- therefore empirically have much less than one chance in one hundred of being observed if the underlying transmission is given by the simulated model. These bounds are shown overlayed with the observational bounds in Figures~\ref{sims_pp} and~\ref{sims_pp_2} corresponding to the rare sources model, radiative transfer model, and uniform UVB cases respectively. The raw predictions from the simulations, without post-processing, are all shown in Figure~\ref{sims}.

\subsection{The Sherwood Simulation}

\begin{figure*}
\centering
\includegraphics[width=0.8\textwidth]{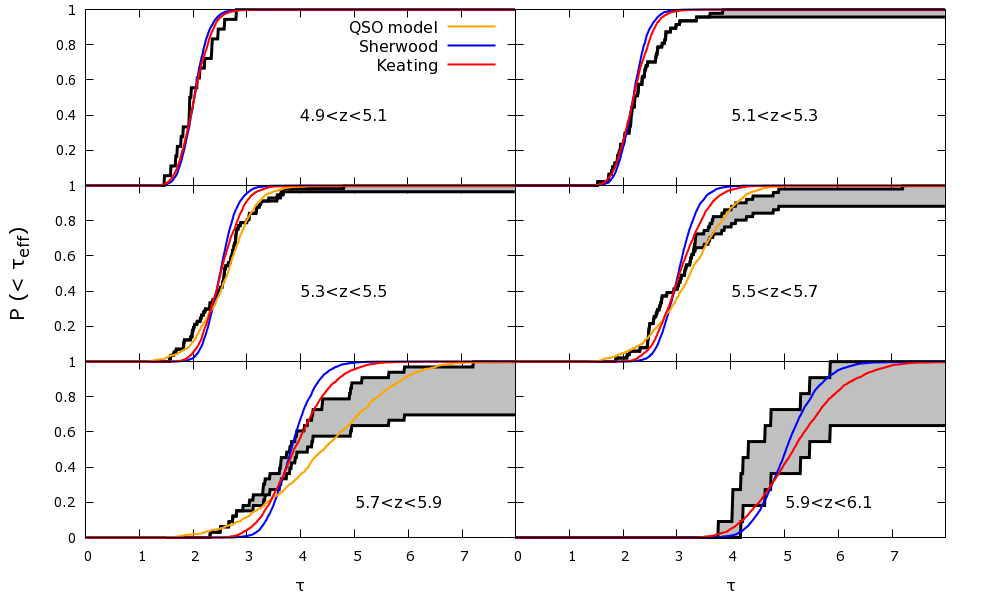}
\caption{Comparison of the measured \lal PDFs at $4.9<z<6.1$ with outputs from a range of numerical simulations. This plots shows only the solid lines from Figures~\ref{sims_pp} and Fig~\ref{sims_pp_2}, and the errors have been omitted for the sake of comparison.}
\label{sims}
\end{figure*}

The Sherwood simulation suite \citep{Bolton17} was designed to reproduce the \lal forest over $2<z<5$ and test its sensitivity to a range of model parameters such as galactic and AGN-driven outflows, thermal histories, and cold/warm dark matter. With gas particle masses of $M_\text{gas} = 9.97 \cdot 10^4 M_\odot$ and $2\times 2048^3$ particles for a box size of 40 cMpc $\text{h}^{-1}$, the Sherwood simulation used here possesses higher mass resolution than other larger-scale cosmological simulations such as Illustris and EAGLE. The simulation is run using the hydrodynamics code P-GADGET 3 \citep{Springel05} and used $\sim2680000$ core-hours of computing time. \citet{Bolton17} compare simulated \lal lines of sight with the results \red{from \citet{Becker13} and \citet{Becker15}}, finding remarkable agreement over $2.5<z<5$ with a slight lack of strongly opaque regions at $z=2.0$. Because it was designed to match observations at redshifts less than $z=5$, the Sherwood simulation uses a uniform UVB with a shape given by the HM12 model, in a scenario where reionisation is driven mostly by galaxies and with no radiative transfer. The interrelations of hydrogen neutral fraction, temperature, photon mean free path, and density are therefore not taken into account, and the simulations are expected to fail in the hydrogen reionisation regime where the UVB is known to be inhomogeneous.

Here we compare the 40 cMpc $\text{h}^{-1}$ Sherwood simulation box with the \lal CDF over $4.9<z<6.1$. Each measurement of $\tau_\text{eff}$ requires stitching together two simulated lines of sight and truncating to 50 cMpc $\text{h}^{-1}$; 2500 total values of $\tau_\text{eff}$ are obtained. The model predictably falls short of reproducing the variety of line-of-sight opacities at $5.3<z<6.1$.
\red{A KS test shows the post-processed sets of predictions are excluded at $5.3<z<5.7$ ($p<0.02$) and marginally disfavoured at $5.7<z<5.9$ ($p<0.1$)}. 
It is perhaps more surprising that the uniform UVB model also fails at $5.1<z<5.3$ ($p<0.03$), where line-of-sight variation is slightly more pronounced than previously reported. Another point of disagreement not reflected by statistical disagreement is that the simulations are never able to output lines of sight with $\tau_\text{eff} > 3.5$ at $z\sim5.2$ - such consistently opaque lines of sight simply do not exist inside the simulated boxes.

\subsection{Radiative transfer simulation}

In addition to the homogeneous-UVB Sherwood simulations, we compare our results to the full cosmological radiative transfer simulations of \citet{Keating17}. 
Following \citet{Daloisio15}, these simulations test the effect of regions ionising at different redshifts on
the spatial variations in the temperature-density relation. The temperature dependence of recombinations rates could then lead to increased fluctuations in \lal opacity. Unlike previous models, Keating et al.'s simulations use an extended and self--consistent reionisation history to boost the IGM temperature. The injected energy and history are chosen to match the temperature and photo-ionisation rates of the IGM at $z\lesssim5.0$ measured using the \lal forest.

This difference to previous models turns out to be important, as \citet{Keating17} find their more realistic choices of reionisation heating do not reproduce the large \lal opacity fluctuations previously reported. Choosing a higher ionisation energy to match the IGM heating in \citet{Daloisio15} they find the fluctuations are still not large enough, and moreover the produced lines of sight are in tension with low redshift \lal forest data as well as transmission peak statistics at high redshift.

The simulations are also run in P-GADGET 3 but snapshots of the resulting density field are then post-processed to include the effects of temperature and ionisation using mono-frequency radiative transfer. Here we use a 40 cMpc $\text{h}^{-1}$ box with $512^3$ mass particles, which should better capture large-scale variations of opacity than the higher resolution 20 cMpc $\text{h}^{-1}$ box. \citet{Keating17} compare their \lal PDFs to data over $4.9<z<5.9$ and find no trend with box size. Here we stitch together two simulated lines of sight for each of the 2500 total measurements of $\tau_\text{eff}$.
We find that this model, while performing better than the Sherwood simulation, does not provide a satisfactory fit to the \lal CDFs. 
\red{Statistical agreement between post-processed lines of sight and observations is improved at all redshifts with only the range $5.3<z<5.7$ being excluded ($p<0.05$). Despite being marginally consistent at lower redshift ($p<0.5$ at $5.1<z<5.3$), the model still does not contain any lines of sight with $\tau_\text{eff}>3.5$. }
Although including radiative transfer does increase line-of-sight variance, the effect is too small by at least factor of two. This is in agreement with the results of \citet{Keating17}.

\subsection{Rare sources simulation}

Finally, we compare the results to predictions of a quasar-driven reionisation toy model from \red{\citet{Chardin17}}. These models include a density field produced by the hydrodynamics code RAMSES onto which radiative transfer is added in post-processing with the ATON code. They found the fluctuations in the UV produced by galaxies in the redshift range $5<z<6$ to be on scales too small ($<50$ cMpc) to account for the spread in \lal opacity. A much better fit was found when rare, bright ionising sources were added to the simulation with a carefully chosen spatial density. Such sources could be (faint) quasar , or alternatively extremely bright star-forming galaxies.

The simulation boxes of \red{\citet{Chardin17}} are only 20 cMcp $\text{h}^{-1}$ in size, but this is not a problem as the line of sight variance is mostly driven by the presence or absence of a strong ionising source nearby. The simulations use $512^3$ mass particles. Because of the smaller simulation volume, we have to stitch three simulated lines of sight together for each measurement of $\tau_\text{eff}$. This results in 1020 measurements, a smaller number than in the previous cosmological simulations.

We find that the rare sources simulations marginally reproduce the CDF at $5.5<z<5.9$, and are the only model to do so out of the ones we tested ($p<0.5$). We note however that the line of sight variance seems to disappear very quickly at lower redshifts, resulting in the model under-estimating the opacity variance at $5.3<z<5.5$. Unfortunately no snapshots were available at $5.1<z<5.3$, but it is unlikely that those would show a sufficient amount of opaque lines of sight since those are already gone from the simulations at $z\sim5.4$. However, we note that there might be room for reionisation to occur later within this model, which could potentially ease the conflict. \red{\citet{Chardin17}} also note that their simulations are in better agreement with the \lal CDFs calculated with smaller window sizes of $l=10$ cMpc~h${}^{-1}$. This is still the case with the updated measurements, as can readily be seen by comparing Figures~\ref{C} and \ref{sims}.

\begin{table*}
\centering
\begin{tabular}{c | c c | c c | c c | c c}
$\Delta z$ & $\overline{F}_\text{fit}$ & $s_\text{fit}$ & $\overline{F}_\text{Chardin}$ & $s_\text{Chardin}$  & $\overline{F}_\text{Keating}$ & $s_\text{Keating}$ & $\overline{F}_\text{Sherwood}$ & $s_\text{Sherwood}$ \\
\hline
4.9 -- 5.1 & $0.146^{+0.023}_{-0.014}$ & $1.15^{+0.30}_{-0.15} $&N/A & N/A & 0.138 & 0.84 & 0.141 & 0.67 \\
5.1 -- 5.3 & $0.112^{+0.007}_{-0.006}$ & $1.34^{+0.20}_{-0.13} $&N/A & N/A & 0.110 & 1.07 & 0.105 & 0.87 \\
5.3 -- 5.5 & $0.094^{+0.012}_{-0.009}$ & $2.21^{+0.33}_{-0.27} $&0.089 & 1.82 & 0.087 & 0.80 & 0.086 & 0.67 \\
5.5 -- 5.7 & $0.063^{+0.016}_{-0.010}$ & $3.6^{+2.4}_{-1.3} $&0.062 & 3.02 & 0.057& 1.50 & 0.057 & 1.00 \\
5.7 -- 5.9 & $0.036^{+0.010}_{-0.005}$ & $4.5^{+3.0}_{-1.2} $&0.038 & 5.70 & 0.035 & 1.82 & 0.030 & 1.44 \\
5.9 -- 6.1 & $0.007^{+0.003}_{-0.002}$ & $2.6^{+2.7}_{-0.9} $&N/A & N/A & 0.007 & 2.80 & 0.005 & 1.78 \\
\hline
\end{tabular}   
\caption{Most likely intrinsic parameter values with 68\% credible intervals, following Equations~\ref{lognorm} and~\ref{params}. Best fit values are marginalised over the other parameter. The errors on simulations are approximately $\pm0.033$ for $s$ and $\pm 0.0023$ for $\overline{F}$, corresponding to the resolution down to which the bayesian likelihood analysis was run. Although the emissivity in the simulations is tuned to match $\overline{F}$, small differences emerge due to the random nature of the forward modelling. The values listed for $\overline{F}_\text{fit}$ here are the values derived from the maximum likelihood fit to the log-normal distribution, not the ones extracted directly from the $\tau_\text{eff}$ measurements, which are more robust and given in Table~4.} \label{table}
\end{table*}

There is also tension between this model and observations for low-opacity lines of sight at $z\sim5.8$: in a QSO-driven scenario, low opacity regions arise purely due to proximity to a source quasar and are therefore not expected to disappear completely at high redshift, as long as the QSOs are active. This appears to be in contrast with observations which do report that transparent lines of sight go missing at $z>5.7$. However, this problem is mitigated by \begin{enumerate} 
\item the smaller observed sample size at these redshifts, which means the discovery of even 1 transparent line of sight could ease the conflict, and 
\item while the ionising emissivity is already tuned to observations, it might be sensible to rescale the predicted opacity by a larger factor to improve the agreement with data (see later).
\end{enumerate}

We note that the volume density of rare bright sources in this model was explicitly chosen to reproduce the \lal CDFs of \citet{Becker15} which use `optimistic' measurements i.e. following our leftmost CDF contour. It is conceivable that the model could be modified to reproduce a CDF closer to the mid-point of our contours. This will be discussed in more detail in the next section.

\begin{figure*}
\centering
\includegraphics[width=0.8\textwidth]{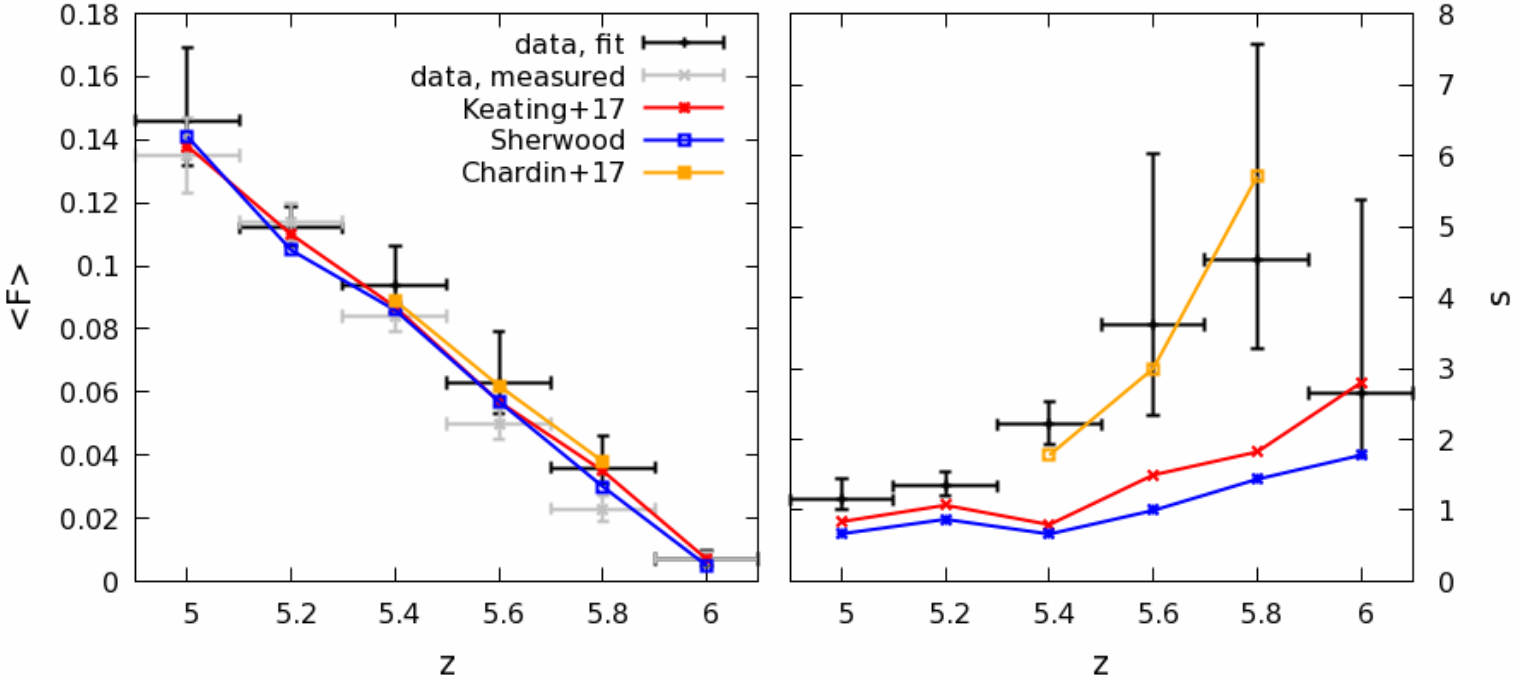}
\caption{Evolution with redshift of the mean flux $\overline{F}$ (\textit{right}) and the skewness parameter $s$ (\textit{right}) of \lal transmission. A rapid decrease in mean flux  with increasing redshift is accompanied by an increase in skewness, as the distribution of fluxes is increasingly non-gaussian. Simulations are post-processed to match the mean flux measured from observations; the offset between observations and simulations at the highest redshift bins reflects the fact that the most likely intrinsic mean flux is lower than the measured mean flux.}
\label{sims}
\end{figure*}

\section{Empirical Analytic Fit}

The treatment of non-detections implemented above is conservative, but does not return a best fit or most likely distribution. Such a fit should ideally  incorporate weighting of the observational constraints based on corresponding observational errors in a way where extra data, no matter how large the errors, should not make the fit worse.
In order to achieve this, we perform a fully bayesian maximum likelihood analysis. This requires a parametrisation of the \lal flux distribution; we empirically find that the data is well fit by a scale-dependant lognormal distribution on the form,
\begin{equation}
P(F;\sigma,\mu) = \frac{e^{ -\text{log}\left(F/\mu\right)^2 / 2\sigma^2}}{F \sigma \sqrt{2\pi}}.\label{lognorm}
\end{equation}
The lognormal parameters $(\sigma,\mu)$ can be recast into the more physically meaningful variables of \textit{mean flux} $\overline{F}$ and \textit{skewness parameter} $s$ following
\begin{equation}
\overline{F} = \mu e^{\sigma^2/2}; \hskip1cm s = \left(e^{\sigma^2} +2\right)\sqrt{e^{\sigma^2} -1}.\label{params}
\end{equation}
This parametrization has the advantage of making the tuning used in simulations explicit: $\overline{F}$ in the simulations is chosen, at each redshift, to match the observations. The difference between data and models is then limited to the shape or skewness parameter $s$, on which meaningful constraints can be obtained. A small value of $s\lesssim1$ indicates a roughly normal distribution, while $s\to \infty$ tends to a log distribution.

For each set of parameters $(\sigma,\mu)$, the likelihood $L$ of the observations given those parameters is computed following 
\begin{equation}
L_{\sigma,\mu} = \prod_\text{obs} \int_{F=0}^{1} P(F; \sigma,\mu) D_\text{obs}
\end{equation}
where $D_\text{obs}$ is the normalised probability distribution of a measurement. This is taken to be a gaussian centered on the observed $F$ and with variance equal to the error on $F$. $D_\text{obs}$ represents the probability distribution of the intrinsic flux within a spectrum region and naturally weights the observations according to their observational errors. We implement the prior that the intrinsic flux is necessarily positive by defining $D_\text{obs}$ only for $F\geq0$. In this way, observations in which the mean flux is formally below the detection threshold, and even observations where the mean flux is formally negative, can be used as constraints on the underlying distribution of fluxes.

Following the likelihood ratio confidence bounds methods the best-fit parameters are found where the value of $L$ is highest, with 68\%, 90\% and 95\% credibility regions found where $L / L_\text{max} > 0.6099$, 0.2585 and 0.1465 respectively (see review in e.g. \citealt{Andrae}). This has the useful feature of being insensitive to multiplicative rescaling of $D_\text{obs}$. 
The post-processed predictions from simulations are obtained by forward modelling 100 simulated lines of sight which then go through the same pipeline as the observations. 

The posterior distribution on $\overline{F}$ and $s$, and predictions from post-processed simulations, are shown in Appendix B. The parameters are non-degenerate and the credible regions of parameter space are therefore well-defined. We marginalise over each parameter in turn by collapsing the likelihood matrix and computing new credible interval bounds. Both parameters increase smoothly with redshift, as shown in Figure~15. The average flux decreases steadily with increasing redshift over $4.9<z<5.9$. While the UVB in simulations is explicitly tuned to match the mean flux, the mean flux recovered after post-processing the lines of sight is slightly lower than measured from observations. This most likely reflects a difference in the clustering of transmission in the simulations: if the transmission occurs in isolated spikes, the flux is more likely to go undetected after observational errors are taken into account (see eg. \red{\citealt{Chardin17-spikes}}).


This decrease in average flux with redshift is accompanied by a smooth increase in the skewness of the distribution until $z\sim5.9$, followed by a tentative decrease at $5.9<z<6.1$. This correspond to an increasingly non-normal distribution of transmission at higher redshift. The decrease in skewness in the highest redshift bin reflects the fact that transparent lines of sight are not found at $z>5.9$. However, no current spectrograph is capable of measuring opacities larger than $\tau\sim8$, and using a log-normal form for residual flux distribution becomes increasingly less appropriate. 

The Sherwood simulation and the model of \citet{Keating17} shows a very modest increase with redshift, while only the model from \red{\citet{Chardin17}} is displaying sufficient non-gaussianity. \red{The highest redshift bin $5.9<z<6.1$ is the only one where the Sherwood and \citet{Keating15} simulations match the data, as the excess in skewness seen at $5.3<z<5.9$ in the data seems to be absent.} Table~\ref{table} gives the most likely values for $\overline{F}$ and $s$ at all redshifts and for all post-processed models. The most likely intrinsic distribution of opacities is plotted in Fig~8 as a dashed line.

\section{Discussion}

Our results confirm the long-lasting inhomogeneity of the IGM opacity to \lal after the end of reionisation. This inhomogeneity is seen to persist as late at $z=5.3$, which is perplexing since photon percolation is predicted to have happened by that redshift by all theoretical models. We confirm the increasing scatter of \lal opacity with higher redshift first reported by \citet{Becker15}. In fact, previous studies may have been too optimistic in their treatments of non-detections of \lal flux. With a large enough sample of lines of sight, we found that the distribution of residual fluxes was aptly described by a lognormal distribution and were able to incorporate constraints from non-detections in a fully consistent way. This choice of distribution is purely empirical, as the lognormal is the most common distributing function for variables which are constrained to be positive, such as flux. Using this parametrisation, we extract a linear increase in mean opacity from $\overline{\tau}_{\text{Ly}\alpha}\sim1.8$ at $z=5.0$ to $\overline{\tau}_{\text{Ly}\alpha}\sim3.8$ at $z=6.0$. It is remarkable, but perhaps not surprising, that all the numerical simulations we confronted with the data required a strongly redshift-dependent rescaling of the source emissivity to match this smooth increase in opacity. 

\red{Another perhaps surprising outcome of our analysis is the lack of binning scale dependence (Section 4.2, Fig 10). The CDF of effective opacity computed with binning scales of 30, 50 and 70 cMpc h${}^{-1}$ are consistent with each other according to a two-sided KS test, with only the case with 10 cMpc h${}^{-1}$ showing deviation. Different processes proposed to drive opacity fluctuations act on different scales; generally speaking, the coherence scales of density and temperature fluctuations are shorter than those of UVB fluctuations (e.g. \citealt{Becker15, Davies16, Keating17}). 
The models currently included in our analysis only span box sizes of 20 cMpc h${}^{-1}$ across, which is insufficient for such an analysis at the present time. 
Explicitly testing this scale-dependence however provides a promising avenue for future work.}

The main caveat in this study, which is also related to a weak point in current numerical simulations, lies in the difficulty of ensuring we are measuring the opacity of the IGM itself as opposed to intervening DLA systems and other absorbers that are not part of the optically thin IGM. Finding these systems via the accompanying decrement to \lal flux is very difficult at the redshifts studied here given how strongly the \lal forest is already absorbed. We have explicitly removed all systems displaying \cfour absorption in the quasar continuum redward of Lyman-$\alpha$. However, weakly ionised transitions such as \magtwo and \oone require far red and infrared spectra of reasonably high resolution and SNR, which we lack in many cases. Systems containing these ions often do not show \cfour in absorption, and their numbers could potentially be increasing at high redshift \citep{Becker06,Bosman17,Codoreanu17}. Nevertheless, one would need to discard the measurements obtained along the most opaque $30\%$ of our sample in order for the remaining lines of sight to match the Sherwood smooth-UVB prediction at $5.1<z<5.3$, and this fraction rises to $60\%$ of the sample at $5.7<z<5.9$. A smooth UVB at those redshifts is therefore confidently ruled out even in the presence of this caveat.
At the same time, recent research has highlighted the crucial importance of including self-shielding effects for numerical simulations of reionisation (e.g. \citealt{Madau17}). Bridging this gap can therefore be done from different angles, as improved surveys will constrain the occurrence rates of low-mass systems in the epoch of reionisation to enable better removal, and numerical simulations become more refined.


\section{Summary}

We have assembled a sample of 62 optical spectra of quasars with $z_\text{source} \geq 5.7$ in order to measure the distribution of IGM \lal transmission over $5 < z < 6$. These objects consist of 13 SDSS-discovered quasars, 10 quasars from DES-VHS, 4 from the SHELLQs survey, 13 from online telescope archives, 19 from previous similar studies and 3 new spectra. The data originates from a total of 10 different optical spectrographs and have been collected over the course of the last 11 years. We use this unprecedented sample of high-$z$ quasars to improve the measurements of residual \lal transmission of \citet{Fan06} and \citet{Becker15}. The large variance in opacity among lines of sight has previously been shown to be incompatible with a uniform UVB. At the same time, unexpectedly large longitudinal correlations in opacity of up to 110 cMpc $\text{h}^{-1}$ at $z>5.5$ mean that only a dramatic increase in number of background sources -- and not a finer sampling -- can aptly quantify cosmic variance \citep{Becker15}. 

We confirm the existence of a long-lasting inhomogeneity in the \lal opacity of the IGM at $5.5<z<5.9$, but also detect a significant departure from the opacity distribution expected for a spatially uniform UVB and temperature-density relation down to $z \sim 5.2$. If the data genuinely reflect large fluctuations in intergalactic opacity at such low redshifts it may present a significant further challenge to models of reionisation the post-reionisation IGM. We also extend our study to $5.9<z<6.1$, finding increased opacity. 

In order to deal with the disparate data quality in our sample and present limits in a transparent way, we then introduce a second bound on the CDF which is ``maximally pessimistic'', i.e. non-detections are taken to mean that $\tau \to \infty$. This allows us to incorporate moderately deep data while being confident that the `true' CDF lies in-between these two bounds; in other words our results present the region permitted by current data.

We further explore an empirical log-normal fit to the CDF, which is characterized by a mean opacity and a skewness. Both detections and non-detections are used to constrain the underlying shape of the opacity distribution. We find a linear increase in mean \lal opacity with redshift from $\overline{\tau}_{\text{Ly}\alpha}\sim2.0$ at $z=5.0$ to $\overline{\tau}_{\text{Ly}\alpha}\sim4.9$ at $z=6.0$, accompanied by a smooth increase in skewness.

Altering the comoving size of the binning window produces only subtle effects on the distribution between the $l = 30, 50 $ and $70$ cMpc $\text{h}^{-1}$ cases at any redshift. A binning with $l = 10$ cMpc $\text{h}^{-1}$ significantly broadens the distribution of effective optical depths at $z<5.7$. 
We also vary the length of the excluded ``proximity region'' which is affected by the source quasar itself, finding no effect at any redshift on the statistical distribution of transmitted flux as long as $\lambda_\text{end, prox} < 1180$ \AA \ is adopted. The traditionally used value of $\lambda_\text{end, prox} = 1176$ \AA \ is thus valid and we don't expect significant contamination from the quasar proximity zones.

We compare our final results with outputs from three different published numerical models: the Sherwood simulation, which uses a spatially uniform UVB \citep{Bolton17}; the radiative transfer post-processed simulation of \citet{Keating17}, which models temperature fluctuations arising from differences in the timing of reionisation; and a model including rare, bright sources from \red{\citet{Chardin17}}. 
Echoing previous works, we find that the data strongly disfavour the uniform UVB model and the radiation transfer model in their current forms. The rare sources model is marginally consistent with data at $z>5.7$ but not at $5.5<z<5.7$. More work may be required to determine whether variations of these models may be more consistent with the present data.

In light of these results, the extreme scatter of \lal opacity at the tail end of
reionisation remains a perplexing puzzle. 


 
\section*{Acknowledgements}

The authors thank the anonymous reviewer for useful and pertinent comments pointing out multiple avenues for future work. SEIB thanks Alberto Rorai and Tricia Larsen for many productive discussions. The authors are grateful to Laura Keating, Jonathan Chardin, and Girish Kulkarni for kindly sharing the outputs of the radiative transfer, rare sources, and Sherwood simulations, respectively. The authors also thank Ian McGreer for agreeing to share reduced spectra from his 2015 paper.

This research has made use of the Keck Observatory Archive (KOA), which is operated by the W. M. Keck Observatory and the NASA Exoplanet Science Institute (NExScI), under contract with the National Aeronautics and Space Administration. Based on data obtained from the ESO Science Archive Facility.  This work is based
in part on observations made with ESO Telescopes at the La Silla
Paranal Observatory under program IDs 084.A-0390, 096.A-0095 and 096.A-0411.

This paper used data obtained with the MODS spectrographs built with
funding from NSF grant AST-9987045 and the NSF Telescope System
Instrumentation Program (TSIP), with additional funds from the Ohio
Board of Regents and the Ohio State University Office of Research. 

Some of the data presented herein were obtained at the W. M. Keck Observatory, which is operated as a scientific partnership among the California Institute of Technology, the University of California and the National Aeronautics and Space Administration. The Observatory was made possible by the generous financial support of the W. M. Keck Foundation. 
The authors wish to recognize and acknowledge the very significant cultural role and reverence that the summit of Maunakea has always had within the indigenous Hawaiian community.  We are most fortunate to have the opportunity to conduct observations from this mountain.

SEIB acknowledges support from the European Research Council Advanced Grant FP7/669253 as well as a Graduate
Studentship from the Science and Technology Funding
Coucil (STFC). Support
by the ERC Advanced Grant Emergence – 32056 is
gratefully acknowledged. XF acknowledges support from the U.S. NSF grant AST 15-15115. LJ acknowledges support from the National Key R\&D Program of China (2016YFA0400703)
and from the National Science Foundation of China (grant 11533001). GB acknowledges support from the NSF under grant AST-1615814.

\red{The Sherwood simulation was performed with supercomputer time awarded by the Partnership for Advanced Computing in Europe (PRACE) 8th call. This project also made use of the DiRAC High Performance Computing System (HPCS) and the COSMOS shared memory service at the University of Cambridge. These are operated on behalf of the Science and Technology Facilities Council (STFC) DiRAC HPC facility. This equipment is funded by BIS National E-infrastructure capital grant ST/J005673/1 and STFC grants ST/H008586/1, ST/K00333X/1.}

\appendix

\section{Mosaic of quasar catalog}

In Figure A1 and A2 we plot all the $z>5.7$ quasar spectra used in this work in a common format. The spectra are normalised by dividing the flux by a best fit power-law. Wavelengths are divided by $z_\text{source} + 1$ to bring the spectra into the rest frame. The y axis is calibrated so that it spans the range $0 \to 5 \times$continuum for each quasar. We do not bin the spectra in order to reflect the diversity of data qualities present in the sample. Error arrays are shown in red.

\begin{figure*}
\centering
\includegraphics[width=0.8\textwidth]{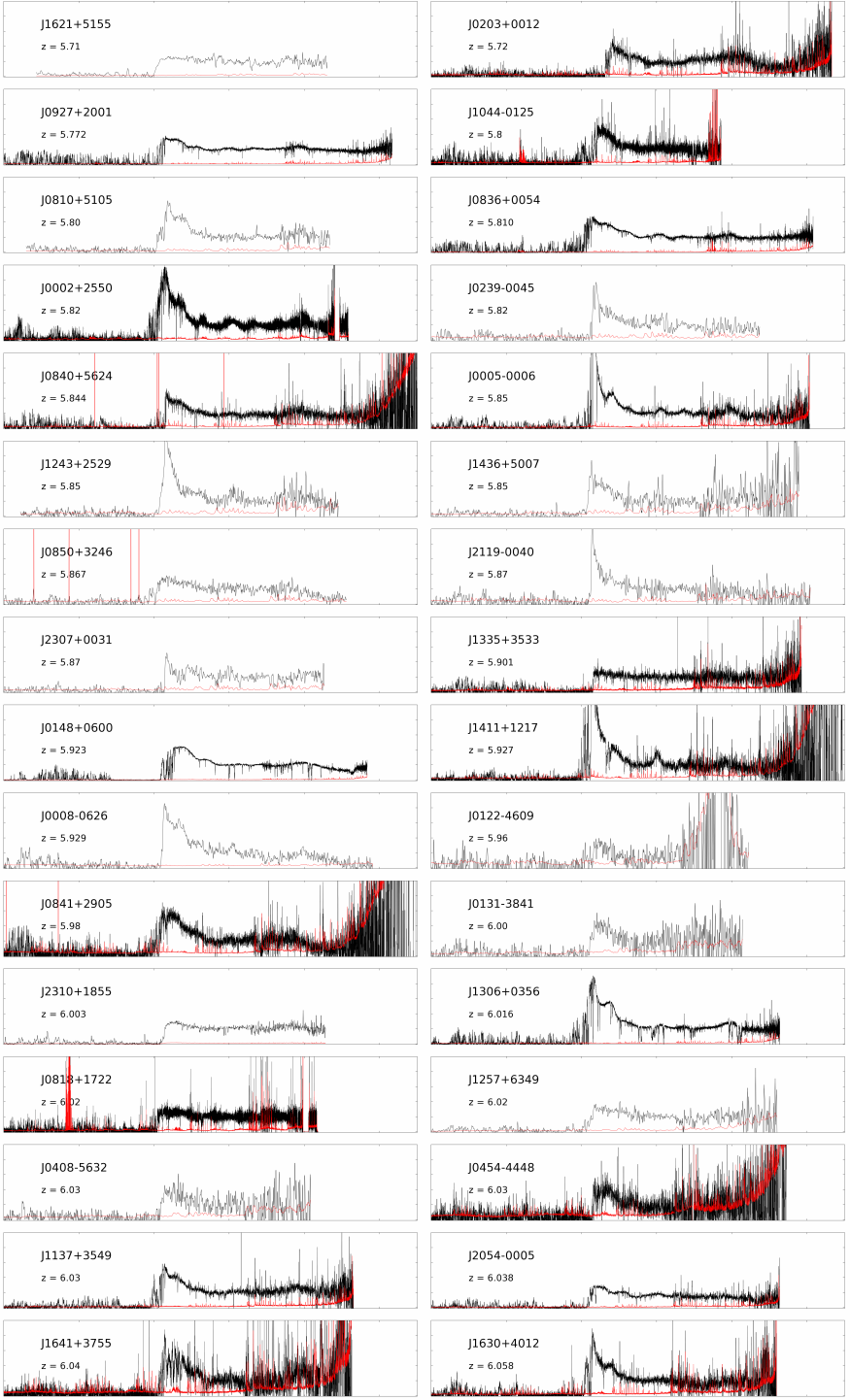}
\caption{First half of the quasar catalog. The origin of each spectrum and the instruments used are listed in Table~\ref{masterfile}. Wavelength runs from 1000 to 1550 \AA \ and the fluxes have been normalised by dividing by the best-fit power law to the continuum.}
\label{fig::mosaicp1}
\end{figure*}
\begin{figure*}
\centering
\includegraphics[width=0.8\textwidth]{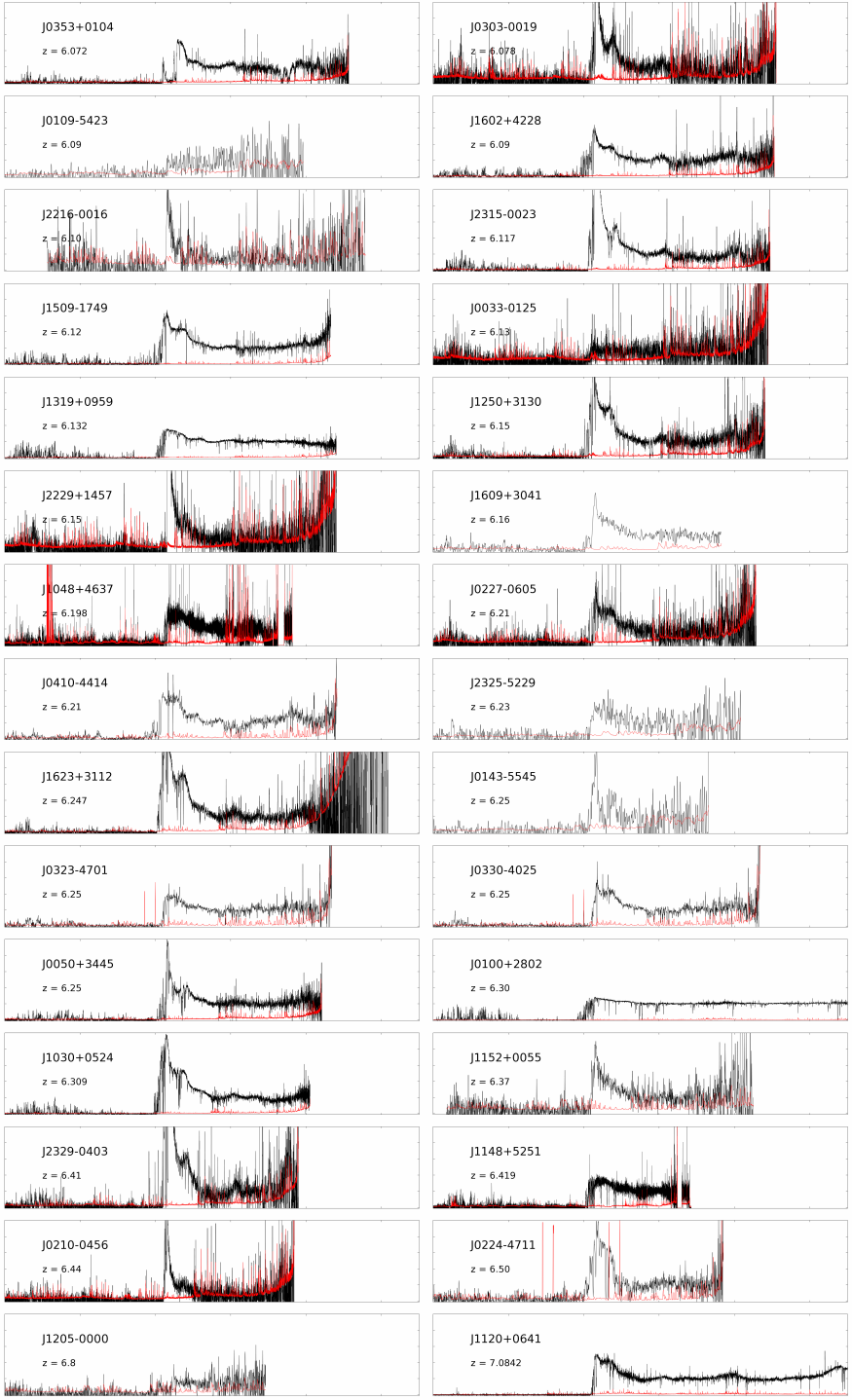}
\caption{Second half of the quasar catalog. Data is as in Fig~\ref{fig::mosaicp1}.}
\end{figure*}

\section{Posterior distribution of $\overline{F}$ and $s$}
In Figure B1 we show the posterior distribution in $\overline{F}$--$s$ parameter space for both the data and best fit values for the post-processed simulations. The results given throughout the paper are marginalised over either one of the two parameters. We plot the skewness on a logarithmic scale to emphasize the fact that the distribution is consistent with being `maximally skewed' at $z>5.9$, which following Equation~\ref{lognorm} corresponds to a exponential distribution tending to infinity at $F=0$. The 68\% and 90\% credible intervals are shown as concentric contours. The colored thick lines correspond to simulations from \red{\citet{Chardin17}} (\textit{orange}), \citet{Keating17} (\textit{red}) and \citet{Bolton17} (\textit{blue}), post-processed as described in Section 5.1.

\begin{figure*}
\centering
\includegraphics[width=0.6\textwidth]{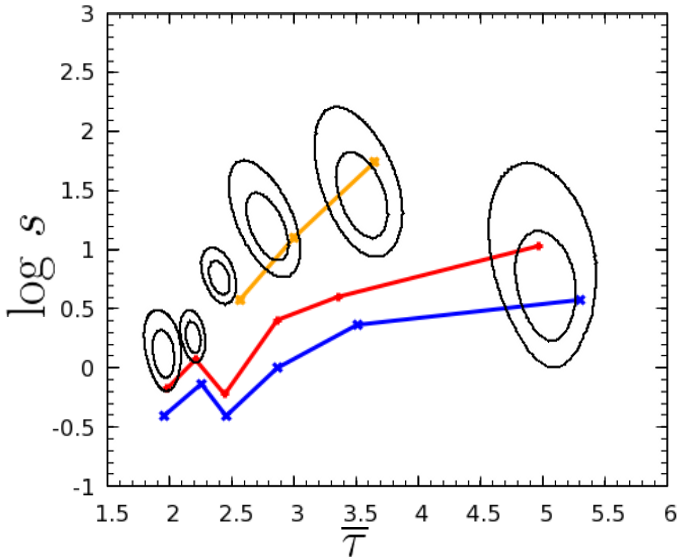}
\caption{Posterior distributions on the skewness $s$ and mean opacity $\overline{\tau} = -\text{log} \ \overline{F}$ of \lal transmission. Different contours correspond to redshift ranges of $\Delta z = 0.2$ beginning at $z=4.9,5.1,5.3,5.5,5.7,5.9$, following the direction indicated by the arrow. The colored thick lines correspond to simulations from \red{\citet{Chardin17}} (\textit{orange}), \citet{Keating17} (\textit{red}) and \citet{Bolton17} (\textit{blue}), post-processed to mimic observational data as described in Section 5.1.}
\label{fig::mosaicp1}
\end{figure*}

\bibliographystyle{apj} \bibliography{bibliography}

\end{document}